**Systematic Assessment of the Static Stress-Triggering Hypothesis using Inter-earthquake Time Statistics**

Shyam Nandan[1], Guy Ouillon[2], Jochen Woessner[3,1], Didier Sornette[4] and Stefan Wiemer[1]

**Affiliations:**

[1]ETH Zürich, Swiss Seismological Service, Sonneggstrasse 5, 8092 Zürich, Switzerland

[2]Lithophyse, 4 rue de l'Ancien Sénat, 06300 Nice, France

[3]Risk Management Solutions Inc., Stampfenbachstrasse 85, 8006 Zürich, Switzerland

[4]ETH Zürich, Department of Management, Technology and Economics, Scheuchzerstrasse 7, 8092 Zürich, Switzerland

**Corresponding Author:**

Shyam Nandan, ETH Zürich, Swiss Seismological Service, Sonneggstrasse 5, 8092 Zürich, Switzerland. (shyam4iiser@gmail.com)

**Key points:**

Compelling evidences for the static triggering hypothesis

Evidence for the static triggering extends to very small Coulomb stress changes

Characteristic time for the loss of the stress change memory is short and independent of the stress amplitude

**Abstract:**

A likely source of earthquake clustering is static stress transfer between individual events. Previous attempts to quantify the role of static stress for earthquake triggering generally considered only the stress changes caused by large events, and often discarded data uncertainties. We conducted a robust two-fold empirical test of the static stress change hypothesis by accounting for all events of magnitude M≥ 2.5 and their location and focal mechanism uncertainties provided by catalogs for Southern California between 1981 and 2010, first after resolving the focal plane ambiguity and second after randomly choosing one of the two nodal planes. For both cases, we find compelling evidence supporting the static triggering with stronger evidence after resolving the focal plane ambiguity above significantly small (about 10 Pa) but consistently observed stress thresholds. The evidence for the static triggering hypothesis is robust with respect to the choice of the friction coefficient, Skempton's coefficient and magnitude threshold. Weak correlations between the Coulomb Index (fraction of earthquakes that received positive Coulomb stress change) and the coefficient of friction indicate that the role of normal stress in triggering is rather limited. Last but not the least, we determined that the characteristic time for the loss of the stress change memory of a single event is nearly independent of the amplitude of the Coulomb stress change and varies between ~95 and ~180 days implying that forecasts based on static stress changes will have poor predictive skills beyond times that are larger than a few hundred days on average.

# 1. Introduction

Earthquakes are thought to interact with each other and alter the times and locations of otherwise inevitable failures by modifying the state of stress at respective locations. Understanding the physics of earthquake interaction may thus provide a path towards explaining the well observed spatio-temporal clustering of earthquakes.

Several mechanisms have been proposed by which the earthquakes can modify the stress on the pre-existing faults. Some of the most notable ones are: (i) static stress change, which is caused by permanent deformation in the vicinity of an earthquake source [i.e. King et al., 1994; Steacy et al., 2005]; (ii) dynamic stress change, which is caused by the passage of seismic waves following an event [i.e. Kilb et al., 2000; Felzer et al., 2006]; (iii) viscoelastic relaxation, which is caused by viscous flow in the lower crust or upper mantle after a moderate to large earthquake [i.e. Freed et al., 2001] and (iv) release of fluids during and after faulting and fluid pore diffusion [Sibson et al., 1975; Sibson, 1982; Hickman et al., 1995]. Moreover, triggering of an earthquake by an antecedent one does not necessarily have to be direct and can also follow an indirect route [i.e. Hill et al., 2002; Helmstetter and Sornette, 2002; 2003; Felzer, 2003; Miller et al., 2004].

In the presence of multiple earthquake interaction mechanisms, it is important to understand their relative significance and to dismiss those that are not supported by data. In this paper we focus only on systematically testing the static stress change hypothesis motivated by is its unique prediction of stress shadow regions, i.e. regions in which seismicity is abated following an earthquake [e.g. Bhloscaidh and McCloskey, 2014].

A wide range of phenomenological observations from various case studies support the static triggering hypothesis. These include studies that show the increase of seismicity rate following moderate to large earthquake in areas that received positive Coulomb stress changes [King et al., 1994; Stein, 1999; Stein et al., 1983; Oppenheimer et al., 1988, Parsons et al., 2000; Hardebeck et al., 1998; Wyss and Wiemer, 2000] and that found consistent observations of stress shadow regions [Kenner and Segall, 1999; Pollitz et al., 2004]. Furthermore, several studies have evaluated the failure probabilities following a static stress change [Stein and Barka, 1997, Hardebeck, 2004, Gomberg, 2005a; Parsons et al., 2000] and their evolution in time using the rate-and-state model [Dieterich, 1994].

On the other hand, several studies have found strong evidences against the static triggering hypothesis. Specifically, Felzer and Brodsky [2005] have shown that stress shadows either do not exist or cannot be distinguished reliably. Marsan [2003] showed that a decrease in seismicity rate following a main shock is very rarely observed in the first 100 days following the main shock. Moreover, the predictive skill of models based on Coulomb stress changes remains poor [Felzer, 2003; Woessner et al., 2011]. In particular, Woessner et al. [2011] found that forecasts of seismicity based on Coulomb stress change tend to be inferior to statistical seismicity forecasts, according to the metrics of the Collaboratory for the Study of Earthquake predictability (CSEP).

Most of the studies focused on the Coulomb stress change caused by specific moderate to large earthquakes and completely ignored the secondary static stress changes caused by aftershocks. Recently, Meier et al. [2014] investigated the role of secondary static stress triggering during the 1992 Landers earthquakes sequence by comparing the triggering potential of cumulative Coulomb stress changes including either main shock Coulomb stress changes caused by moderate to large earthquakes (M>6) or secondary Coulomb stress changes caused by only smaller earthquakes (2<M<6). They defined the triggering potential in terms of Coulomb index, defined as the fraction of the total number of earthquakes that received net positive Coulomb stress change by the time of their occurrence. They found that by including the secondary stress changes the Coulomb index dropped to 0.79 from 0.85, which was obtained by only considering main shock Coulomb stress changes. The authors attributed this slight drop in Coulomb index to large uncertainties in secondary Coulomb stress changes as for example the uncertainties in the source parameters of small earthquakes. However, Helmstetter [2003] and Marsan [2005] have found empirical evidence that small earthquakes could play an equally or even more important role in earthquake triggering within the framework of the Epidemic Type Aftershock Sequence (ETAS) model. Augmenting further support for the secondary stress changes, Felzer [2003] showed that aftershock probability maps based on using only times and locations of previous aftershocks while completely ignoring the main shock-induced stress changes can outperform the forecasts made using only the main shock Coulomb stress changes. Taking account of secondary stress changes can also help explain why a significant fraction of aftershocks occur in stress shadow regions, as the secondary aftershocks of a main shock are not physically constrained to only occur in regions where the stress change caused by the main shock was positive [Felzer, 2002].

Another indispensable consideration for testing Coulomb stress changes is the choice of fault planes on which Coulomb stress changes are resolved. In the past, researchers have generally adopted two approaches. The first one consists in resolving the Coulomb stress changes on fault planes that are optimally oriented for Coulomb failure [as in Stein et al., 1992; King et al., 1994], which are determined based on the information of magnitude and direction of the principal axes of the regional stress field. The second approach consists in using the information of the mapped fault network [Steacy et al., 2005; McCloskey et al., 2003; Bhloscaidh and McCloskey, 2014]. Arguments exist for and against both approaches. Proponents of the former approach argue that the later can only be applied to very limited well-documented faults while ignoring the blind faults that can present major threat. On the other hand, proponents of the later approach argue that the former relies on the existence of optimally oriented fault planes everywhere in the crust despite mounting observations against it [Steacy et al., 2005]. Moreover, one needs to know the poorly constrained regional stress field a priori for finding optimally oriented fault planes. Several other researchers [e.g. Hardebeck et al., 1998; Steacy et al., 2004; Meier et al., 2014] have computed Coulomb stress changes using the focal mechanisms of earthquakes. However, they are confronted with the focal plane ambiguity, which is often dealt with by making a random choice between the two nodal planes. This adds an extra layer of uncertainty to already uncertain Coulomb stress changes and can possibly obscure one's resolution to accept or reject the static stress change hypothesis.

Model and data uncertainties while evaluating Coulomb stress changes are, with some exceptions [e.g. Hainzl et al., 2009; Woessner et al., 2012; Catalli et al., 2013; Cattania et al., 2014], rarely considered when evaluating the static stress change hypothesis, which can lead to false acceptance or rejection.

In this study, we test the static stress hypothesis using all earthquakes with magnitude equal to and larger than 2.5 listed in the focal mechanism catalog of Southern California [Yang et al., 2012] as our primary dataset. For each event, we first solve the focal plane (and slip vector) ambiguity. This is done using the clusters of earthquakes defined in the relocated catalog of Southern California [Hauksson et al., 2012]. The clusters are assumed to reveal the fault planes on which the earthquakes occur. We then compute the Coulomb stress change interaction between all causal source-target pairs. In contrast to previous stress change studies, all earthquakes are considered as the source of Coulomb stress change at the location of subsequent earthquakes including location and focal mechanism uncertainties while resolving the focal plane ambiguity and evaluating the Coulomb stress change. Following the evaluation of Coulomb stress change, we address the static stress change hypothesis from two perspectives. First, we investigate how the time variation of the seismicity rate depends on the sign and amplitude of Coulomb stress changes. We fit the rate of triggered events by an exponentially tapered Omori law, superimposed over a constant background rate, for different amplitudes of Coulomb stress changes. We then use the best-fit parameters for different stress bins to evaluate the hypothesis. Second, we analyze the same dependency for the Coulomb Index (CI), the fraction of events that received net positive Coulomb stress changes compared to the total number of events. All CI values are then compared to a Mean-Field CI, i.e. an expected average value, derived from the time-independent structure of the fault network.

In section 2, we describe the data that is used for the analysis. Section 3.1 describes the choices for various parameters used for computation of Coulomb stress changes. Section 3.2 and 3.3 describe our analyses in details. Section 4 presents our various results. In section 5, we interpret the results and explain their implications. Section 6 summarizes our conclusions. For the convenience of the reader, we have defined the frequently used acronyms and symbols in Table 1 in the order of their appearance in the paper. In the interest of length, we have described several important analyses and corresponding results related to the present work in the accompanying supplementary material.

**2. Data**

**2.1 Earthquake data: Threshold magnitude, focal mechanisms and locations**

We primarily use the focal mechanism catalog of southern California (YANG) covering the period 1981-2010 [Yang et al., 2012] that includes location, focal mechanism and corresponding uncertainties in the focal mechanisms of 179,255 events. As fault plane solutions for source and receiver are necessary for computing Coulomb stress change ($\Delta CFS$), the catalog provides an excellent opportunity to analyze the role of small earthquakes in the stress redistribution process. The YANG-catalog does not provide the uncertainty in earthquake locations. So, we also use the relocated catalog (HAUK) of southern California, which covers the period 1981-2011 [Hauksson et al., 2012] to assign uncertainties to the location of earthquakes present in the YANG-catalog. This is simply done using the event-id information present in both catalogs. Note that the YANG-catalog is a subset of the HAUK-catalog, thus all earthquakes in the YANG-catalog can be assigned a location uncertainty.

The importance of the HAUK-catalog for our analysis is the clustering information of earthquakes it features. The clusters are defined on the basis of earthquake waveform similarity and we use this information to resolve the focal plane ambiguity in the YANG-catalog (see section 3.2).

For evaluating the Coulomb stress hypothesis, there is no clear reason or evidence to accentuate the importance of using a complete catalog. However, one can argue that, if the Coulomb stress hypothesis is correct, then we would expect triggering to initiate earlier in the regions where main shocks cause positive Coulomb stress changes compared to the regions where they cast a stress shadow. However, just as a consequence of proportion, there would be more missing events in a positive stress change lobe than in a negative one. This can bias the outcome of the test against the Coulomb stress change hypothesis. So, it is essential to estimate a magnitude threshold above which the catalog is approximately complete. However, for simplicity, we only consider a space-time independent magnitude threshold for further analysis, similarly to [Steacy et al., 2004, Meier et al., 2014]. Nevertheless, in Text S2, we explore the effect of alternative magnitude thresholds on our results.

Both, the YANG- and the HAUK-catalog, follow the Gutenberg-Richter (GR) law for magnitudes larger than $m_t = 2.2$ and $m_t = 2.5$, with b-values of 0.97 and 1, respectively (Figure S1). We have determined $m_t$ and b-values using the method proposed by Clauset et al. [2009]. We use $m_t = 2.5$ for the YANG-catalog on the basis that it cannot be more complete than its parent HAUK-catalog. This reduces the total number of usable events to 21,480 earthquakes for the Coulomb modeling procedure. We observe that the YANG-catalog misses events across the entire magnitude range up to M=6.2 and is incomplete in the strictest sense compared to the HAUK-catalog. Above $m_t = 2.5$, the latter catalog contains 17,856 earthquakes more than the former, with most of the missing earthquakes in the YANG-catalog located in offshore regions and Mexico, which have poor station coverage.

## 2.2 Finite-fault source models

Coulomb stress changes strongly depend on the details of the slip distribution on finite faults in the "near-field" of the source[Woessner et al., 2012]. Case studies such as that of Steacy et al. [2004] underlined the importance of choosing slip solutions incorporating correct rupture geometry over simple slip solutions based on empirical relations and focal mechanism by comparing the consistency of off fault aftershock distribution with Coulomb stress change caused by main shock in both cases. Unfortunately, detailed slip inversions are rarely available for small earthquakes. Detailed slip distributions are available for only 10 large earthquakes present in the YANG-catalog from the online Finite-Fault Source Model Database (http://equake-rc.info/srcmod/). The names of these earthquakes for which the detailed slip models are available are listed in Table 2 along with the relevant references. There are generally multiple solutions available for each of these earthquakes. We do not prefer any of these slip models and treat them as epistemic uncertainty of the true source slip distribution. This uncertainty results from multiple inversion procedures, data used for the slip inversion, uncertainty in the data and so on. Thus, we shall randomly choose between the slip models for those earthquakes.

## 3. Method

### 3.1 Coulomb Stress computations

#### 3.1.1 Method and parameter values

We compute the Coulomb stress changes using the code of Wang et al. [2006], which uses the solutions of Okada [1992] for internal displacement and strains due to shear and tensile faults in a homogenous half-space for finite rectangular sources. We assume a constant shear modulus $\mu_B$=32 GPa, a coefficient of friction $\mu = 0.6$ and a Skempton's Coefficient B=0.75 for most of this study. The value of $\mu$ is chosen on the basis of empirical evidences from Byerlee [1978] for most rock types. The Skempton's coefficient B is usually found to vary between 0.5 and 1 [Cocco, 2002; Green and Wang, 1986; Hart and Wang, 1995]. Various studies have found that these two parameters have only modest effect on the aftershock correlations with $\Delta CFS$ [e.g. King et al., 1994, Catalli et al., 2013]. However, for the sake of completeness we also explore the effect of alternative values of $\mu$ and B on our results in (Text S2).

### 3.1.2 Simple Source models

The computation of $\Delta CFS$ requires a specification of the size of, and slip on, the source fault. We assume a homogenous slip distribution on rectangular faults for all the earthquakes for which detailed slip distribution models are not available. We estimate the size of the faults (length and width) and the amplitude of the slip given the corresponding magnitude and style of faulting using the empirical relations from Wells and Coppersmith [1994]. These relations have been summarized in Table 3.

We classify the focal mechanisms present in the catalog as strike-slip, normal, thrust and oblique type based on the criteria proposed by Frohlich et al. [1992]. The criteria are that an earthquake is strike slip, normal or thrust type depending on whether $sin^2\delta_B > 0.75$, $sin^2\delta_P > 0.75$ or $sin^2\delta_T > 0.59$ respectively. $\delta_B, \delta_P$ and $\delta_T$ are the angles between the horizontal and the B axis, P axis and T axis of the earthquake focal mechanism. If none of these criteria is fulfilled, the earthquake is classified as oblique-type. Out of 21,480 earthquakes above magnitude 2.5, ~47% are strike slip, ~9% are normal, ~12% are thrust and the rest (~32%) are oblique-type earthquakes.

A uniform slip model is then obtained by assigning to each point on the fault a slip vector with magnitude $\bar{u}$ and direction given by the rake of the preferred nodal plane. The orientation of the fault is given by the strike and dip of the preferred nodal plane, which choice is discussed in the section 3.2.

Several other flavors of slip distribution could be used instead of the uniformly distributed slip that we proposed above. One of them is to use a tapered slip distribution with slip being zero at the edges of the fault. The argument given in favor of such a slip distribution is that it does not lead to strong stress singularities at the edges of the fault. However, Steacy et al. [2004] noted that the percentage of Landers aftershocks that received positive Coulomb stress change from the Landers earthquake on one or both nodal planes is almost the same when using tapered and uniform slip distributions, which suggests that our result should not depend on this choice.

**3.2 Focal Plane Ambiguity**

A focal mechanism defines two nodal planes and in most cases it is unclear which one ruptured. The importance of the choice of the nodal plane for the present study is that the computed Coulomb stress change depends on strike, dip and rake of both source and receiver faults.

In this work, we use the earthquake clusters present in the HAUK-catalog as information to resolve the nodal plane ambiguity (Text S1). These clusters are formed on the basis of waveform cross-correlation (see details in Hauksson et al., 2012), which measures the similarity between two waveforms. A high cross correlation between the waveforms of two earthquakes at a given set of stations implies that the two earthquakes have very similar focal mechanisms and their relative location is such that heterogeneities in the velocity cause very small signal scattering [Waldhauser et al., 2000].

It is tempting to assume that all events present in a cluster occurred on the same planar structure. However, several of the clusters are composed of multiple sub-clusters of earthquakes, which can potentially represent different faults with varying orientations as illustrated with the epicenters of earthquakes belonging to the cluster with unique identification number #50106 (Figure S4). Note that we have created the five-digit unique identification number for each cluster by combining the polygon index (varying from 1-5) in which the earthquakes belonging to the cluster are located, with the similar event cluster identification number (varying from 0001-4317) originally assigned to each cluster. The epicenters indicate the presence of multiple fault segments. Even though the earthquakes seemingly belong to faults with different orientations, their assignment to a single cluster can be understood in the following way. In the HAUK-catalog, two events are designated as similar if the cross correlation of their waveforms yields at least 8 cross correlation coefficients from all stations larger than 0.6, and if the average of the maximum cross-correlation coefficients from all stations is larger than 0.4 [Hauksson et al., 2012]. By increasing or decreasing these preset thresholds, one can further increase or decrease the similarity between the pairs of events, thereby decreasing or increasing the possibility of two events belonging to different faults being assigned to one cluster. Furthermore, earthquakes belonging to conjugate faults would produce approximately similar waveforms at a given station if the path travelled by corresponding waves were approximately the same. This explanation is indeed verified in the cluster shown in Figure S4 where we see the presence of such near-conjugate faults.

In each cluster of the HAUK-catalog, we further need to identify sub-clusters of earthquakes that could be designated as occurring on a single fault segment. We apply a method similar to Ouillon and Sornette [2011] to each cluster of the HAUK-catalog in order to reconstruct the seismically active part of the fault network (Text S1.1-2). These reconstructed faults are then used to resolve the focal plane ambiguity (Text S1.3).

### 3.3 Appraisal of Coulomb stress change hypothesis

### 3.3.1 Deriving waiting-time distributions

Our first approach is to estimate the parameters of the conditional waiting time distribution conditioned on the Coulomb stress change. For clarity of exposition, we have outlined our general methodology in the flow chart shown in Figure S2.

We define the waiting time, $t_{ij}$, between two earthquakes $E_i$ and $E_j$ as:

$$t_{ij} = t_j - t_i \qquad (1)$$

In Equation 1, $t_i$ is the occurrence time of earthquake $E_i$ that causes the $\Delta CFS_{ij}$ Coulomb stress change at the location of earthquake $E_j$ that occurs at time $t_j$. $\Delta CFS_{ij}$ is considered causal if $t_{ij} > 0$ and acausal otherwise.

We assume that a sudden stress change at a given location alters the waiting time to the next earthquake. A positive $\Delta CFS_{ij}$ should shorten $t_{ij}$ while a negative $\Delta CFS_{ij}$ should increase $t_{ij}$ if the static stress change hypothesis is correct. With 21,480 earthquakes in the catalog above $m_t = 2.5$, we have ~230,684,460 causal pairs. Computing more than one hundred million Coulomb stress change interactions is computationally expensive, thus we reduce the computation time by only computing $\Delta CFS_{ij}$ if the ratio of the source-receiver distance to the source event length is less than 10 assuming that the influences at larger distances is negligible:

$$\frac{d_{ij}}{L_i} \leq 10 \qquad (2)$$

Here, $d_{ij}$ is the hypocentral distance between earthquake $E_i$ and $E_j$ and $L_i$ is the rupture length of earthquake $E_i$ (Table 3). Applying this constraint reduces the total number of causal pair interactions to ~2,500,000 pairs, enabling effective result generation and analysis.

We then sort $t_{ij}$ according to the absolute amplitude ($|\Delta CFS_{ij}|$) of the static stress changes and divide the latter into $k$ different stress bins, where $k$ varies between 1 and n$_{bin}$(=50), where n$_{bin}$ is the total number of Coulomb stress bins. Each of these stress bins contains an equal number of waiting times ($N^k \sim 50{,}000$). We consider the median of $|\Delta CFS_{ij}|$ in the k$^{th}$ stress bin, $|\Delta CFS_{ij}|_k^{median}$, as the representative stress value of the k$^{th}$ stress bin. Each of the stress bins contains two groups of waiting times, $t_{ij}^{k+}$ or $t_{ij}^{k-}$, depending on whether the associated Coulomb stress change was positive or negative. The total number of waiting times in any bin, $N^k$, is equal to the sum of number of waiting times for which the associated Coulomb stress change is positive ($N^{k+}$) and negative ($N^{k-}$).

For both groups, we model the probability density function (PDF) of waiting time distribution ($W$) as a tapered Omori-decay using the form:

$$W(t,K,c,p,\tau,B,\theta) = \frac{\left(\frac{K}{(t+c)^p}e^{-t/\tau}+B\right)e^{-t/\theta}}{\int_0^T\left\{\left(\frac{K}{(t+c)^p}e^{-t/\tau}+B\right)e^{-t/\theta}dt\right\}} \qquad (3)$$

Equation 3 respectively consists of a triggering and a background rate component, tapered by an exponential term to model the finiteness of the catalog. The triggering component, $\frac{K}{(t+c)^p}e^{-t/\tau}$, resembles the kernel used for modeling an Omori decay along with an exponential taper with a characteristic time $\tau$ beyond which the rate of aftershocks exponentially decays to the background seismicity rate B. T denotes the length of the catalog.

In a catalog, the sources that occur earlier in time have more available targets than the ones that occur later in time due to the finite size of the catalog, which makes the waiting time distribution biased at long times. To model this bias, we introduce an exponentially decaying term to modulate the whole waiting time distribution, with characteristic time $\theta$, which is a generic way to model finite-size effects.

The parameters of the waiting time distribution, $\lambda = \{K, c, \rho, \tau, B, \theta\}$, are then obtained by maximizing the log-likelihood. Equation 4 gives the log-likelihood of the waiting time distribution.

$$LL = \sum_{m=1}^{n}\{\log(W(t_m, K, c, \rho, \tau, B, \theta))\} \quad (4)$$

Here, $\{t_m, m = 1, \ldots, n\}$ are the observed waiting times in the stress bin for which the log-likelihood is being optimized and n is the total number of waiting times present in that bin. We have numerically maximized LL for all k stress bins to obtain two sets of parameters $\lambda^{k+}$ and $\lambda^{k-}$ and respective maximum log likelihood per data point, $MLL^{k+} = \frac{\max(LL^{k+})}{N^{k+}}$ and $MLL^{k-} = \frac{\max(LL^{k-})}{N^{k-}}$, corresponding to the waiting times $t_{ij}^{k+}$ and $t_{ij}^{k-}$ respectively.

We then define R as the ratio of the instantaneous triggered rates to background rates

$$R = \frac{K}{Bc^\rho} \quad (5)$$

Ideally, the spatial volumes covered by the corresponding stress bins should normalize the seismicity rates in all stress bins. However, computing the volume of different stress bins to normalize the corresponding seismicity rates is not simple. However, since both background seismicity rate and triggered seismicity rate sample the same volume for a given stress bin, taking their ratio removes this effect. Using the two sets of parameters, $\lambda^{k+}$ and $\lambda^{k-}$, we obtain two sets of ratios for the instantaneous triggered rates to background rates, $R^{k+}$ and $R^{k-}$, for each stress bin.

We then investigate the dependence of R on Coulomb stress changes. In particular, we compare the values of $R^{k+}$ to $R^{k-}$ for the k[th] stress bin with the median Coulomb stress change amplitude $|\Delta CFS_{ij}|_k^{median}$.

### 3.3.2 Uncertainties in parameters of the waiting time distribution

We compute uncertainties in the parameters $\lambda^{k+}$ and $\lambda^{k-}$ for all the k stress bins. The uncertainty in these parameters comes from the conditioning of the distribution of $t_{ij}$ on $\Delta CFS_{ij}$. Sources of uncertainties in Coulomb stress changes are many and we consider only data uncertainties as outlined above: location, focal mechanism and choice of the correct nodal plane.

To compute the uncertainty in $\lambda^{k+}$ and $\lambda^{k-}$, we first randomly perturb the location of the earthquakes present in the relocated and focal mechanism catalogs according to the respective absolute horizontal and depth uncertainties. We also perturb the focal mechanisms of earthquakes according to their corresponding uncertainty. We then determine the preferred normal and slip vectors for each earthquake. We make the choice of nodal planes according to this method for all earthquakes except for the 10 earthquakes for which slip models are available. For the latter, we randomly choose between the available slip models. With the perturbed earthquakes locations and preferred normal and slip vectors for each earthquake in the focal mechanism catalog, we compute the estimates of $\lambda^{k+}$ and $\lambda^{k-}$ for the given realization of the catalog. Repeating these steps 1000 times, we obtain 1000 estimations of $\lambda^{k+}$ and $\lambda^{k-}$, which are then used to compute the median value and the 2.5-97.5% confidence intervals of each parameter within each stress bin.

### 3.3.3 Coulomb Indices (CI)

The second metric that we use to validate the static triggering hypothesis is the Coulomb Index (CI), traditionally defined as the fraction of earthquakes in the catalog that received a net positive Coulomb stress change from all the preceding earthquakes [Hardebeck et al., 1998]. However, this definition suffers from two main deficiencies: the first arises from the finiteness of the catalog. The net Coulomb stress change at a receiver event can vary significantly if one varies the time length of the catalog. Secondly, the obvious incompleteness of the focal mechanism catalog influences the result. The largest Coulomb stress contributor to a given target could be a very small event, by the virtue of its spatial proximity, not recorded in the catalog at all.

To avoid these two major deficiencies, we use a modified definition of CI values. We define CI as the fraction of causal pairs in the earthquake catalog with positive Coulomb stress change interaction ($\Delta CFS_{ij}>0$). Note that the causal pairs have to satisfy the distance/length constraint (Equation 2) for the sake of faster computations. In this definition, the CI value is computed from a representative subsample of all causal Coulomb stress changes without bias.

Moreover, unlike previous studies [Hardebeck et al., 1998; Meier et al. 2014] we extend the analysis by categorizing the CI value as a function of Coulomb stress change and waiting times between earthquakes. The CI value for the $k^{th}$ stress bin at time $t$ is then equal to:

$$CI_k(t) = \frac{N^{k+}*W^{k+}(t)}{N^{k-}*W^{k-}(t)+N^{k+}*W^{k+}(t)} \qquad (6)$$

Here, $W^{k+}$ and $W^{k-}$ are the analytical waiting time distributions with parameters $\lambda^{k+}$ and $\lambda^{k-}$ that best model the distribution of $t_{ij}^{k+}$ and $t_{ij}^{k-}$ respectively. $N^{k+}$ and $N^{k-}$ are the total number of source receiver pairs with positive and negative Coulomb stress changes, respectively. Note that $N^{k+} + N^{k-} = N^k$ is constant for all stress bins. The estimate of $CI_k(t)$ does not involve any binning in time as we use the proposed theoretical waiting time distributions with the parameters estimated for different stress bins.

For the static triggering hypothesis to hold, the $CI_k(t)'s$ should be significantly larger than a Mean-Field Coulomb index (MFCI), at least above a certain stress threshold. The definition of the MFCI is extremely important for accepting or rejecting the static stress hypothesis. A suitable choice for the MFCI depends on the geometry of the underlying fault network. As pointed out by Meier et al. [2014], the observed CI value may simply be a consequence of the underlying fault network, which dictates the location and focal mechanism of earthquakes. Only after we decouple the effect of the underlying fault network from the observed CI value, are we able to see the effect of static stress triggering if it exists.

To compute the MFCI, we first randomly shuffle the order of the events in time while keeping each event associated to its original spatial location and preferred nodal plane (see Supplementary Text S1). Hereby, we approximately remove the space-time causality between the events, which might exist due to the static stress triggering. We then compute the Coulomb stress change interactions between causal pairs for this randomly shuffled catalog as for the original catalog. Further, we compute the CI value for each reshuffled catalog by taking the ratio of the total number of positive

Coulomb stress interactions to the total number of pairs. This gives a MFCI that we would observe if the events were not causally related to each other. Note that the MFCI reflects properties of the fault network since the geometry is conserved in the time randomization procedure. On the time scale of the catalog, the fault network is assumed to be static and not evolving. Thus, MFCI should also be independent of the specific realization of the waiting times between all earthquake pairs in the real (non-shuffled) catalog.

### 3.3.4 Uncertainty of the CI

Our computed $CI_k(t)$ features uncertainties rooted in $\Delta CFS_{ij}$ uncertainties. The uncertainty in the $CI_k(t)'s$ is simply the manifestation of the uncertainty in the parameters, $\lambda^{k+}$ and $\lambda^{k-}$, and in the total number of positive and negative Coulomb stress change interactions in a given Coulomb stress change bin, $N^{k+}$ and $N^{k-}$, according to Equation 6. As a result, the uncertainty in $CI_k(t)$ can be simply computed once we have computed the uncertainty in $\lambda^{k+}$, $\lambda^{k-}$, $N^{k+}$ and $N^{k-}$. We also compute the uncertainty in MFCI by computing a MFCI as described in section 3.3.3 for each of 1000 focal mechanism catalogs perturbed according to location and focal mechanism uncertainties.

### 3.3.5 Statistical significance using Wilcoxon Ranksum test

We primarily want to test the statistical significance of $R^{k+}$ and $CI_k(t)$ respectively over $R^{k-}$ and MFCI. This is done using the right tailed Wilcoxon Ranksum test [Wilcoxon, 1945]. Given two populations, A and B, we test the null hypothesis ($H_0$) that the medians of some quantity (X) measured for both populations are equal against the alternative hypothesis ($H_A$) that the median X for population A is larger than that of population B at some predefined significance level, $\alpha$. The null hypothesis is rejected if the p-value or $p_{val}$, which is the probability of obtaining the test statistic at least as extreme as the one that was actually observed (assuming that the null hypothesis is true), is found to be smaller than $\alpha$. A standard value of $\alpha$ used in the literature is 0.05 or 0.01.

For instance, to test the statistical significance of $CI_k(t)$ against the MFCI for the k$^{th}$ stress bin, we test the null hypothesis that, at a given time $t$, the median of $CI_k(t)$ is equal to the median of the mean-field CI value against the alternative hypothesis that the median value of $CI_k(t)$ is larger than the median value of MFCI at significance level of 0.01.

**3.4 Effect of random choice of the nodal plane**

We repeat the analysis described in section 3.3 using a random choice of the nodal plane for each earthquake instead of the preferred nodal plane according to the method proposed in section (Text S1). We then compare the results for the preferred choice of the nodal plane to results obtained with the random choice. This is done to check that stress, and not strain, is the relevant quantity for triggering, and that the knowledge of the underlying network improves our ability to study quantitatively such a process.

**4. Results**

In the following sections, we refer to the case where preferred nodal planes are used as PREF, and the case where randomly chosen nodal plane are used as RAND.

**4.1-Goodness of fit of the tapered Omori-law to empirical waiting time distributions**

As our interpretations about the validity of the static stress change hypothesis rely on the variation of the parameters of the proposed waiting time distribution that should be related in a significant way to the Coulomb stress changes, it is imperative that the proposed waiting time distribution fits the data well. We demonstrate the goodness of fit visually by plotting the empirical and inverted waiting time distributions for one of the positive and negative Coulomb stress bins ($|\Delta CFS_{ij}|_k^{median} \sim 42000 \, Pa$) for PREF and RAND (Figure 1). The different scattered markers represent the empirical waiting time distributions, which are obtained using kernel density estimation with an Epanechnikov kernel [Epanechnikov, 1969] with a smoothing parameter of 0.01 and 1000 time bins, for the four possible cases shown in the legend. The binning in time has been done logarithmically, which implies that the size of the time bins remains constant on logarithmic scale (base 10). Solid lines represent the waiting time distribution whose parameters have been obtained by maximizing the log likelihood for the tapered Omori-law (Section 3.3.1). The optimal parameters corresponding to each case are shown in Table 4. The large scatter in the empirical distributions at small times can be attributed to the very small size of the time bins, which contain fewer events and thus exhibit larger statistical fluctuations. Also note that, the apparent decays in the waiting time distributions at waiting times smaller than $10^{-3}$ days are artifacts. As the size of the time bins become very small, they either contain one or no observation. Since the size of consecutive time bins increases with as constant factor, the resulting empirical distributions obtained by normalization of the number of observations in each time bin by its size show apparent power-law decays with unit exponents.

We identify roughly four regimes in Figure 1 that can be observed in the empirical waiting time distributions, which have been modeled using the proposed form of the waiting time distribution (Equation 3). The "Pre-Omori Regime" corresponds to waiting times smaller than c. In the "Omori Regime", the waiting distributions exhibit a power-law decay with an exponent equal to $\rho$. The power-law decay then transitions to a constant background rate, B, through an exponential taper with a characteristic time $\tau$. Finally, the distribution is subjected to the effects resulting from the finite size of the catalog and decays nearly exponentially for waiting times $> \theta$.

We find that the proposed form of the waiting time distribution with the optimal parameters fit the empirical distribution well except in the upper tail in all cases. In the upper tail, the empirical waiting time distribution seems to decay faster than the proposed form. For improvement of the fit, we might need to modify the exponential taper used to model the finiteness of the catalog to even faster decaying taper. However, the fits in other parts of the distribution are reasonable and we speculate that the values of the maximum likelihood estimates of the parameters except $\theta$ would only change marginally after such a modification.

Finally, in Text S4 we compare the relative goodness of fit of inverted models for all the stress bins and all the cases. We find that the inverted models fit the real data equally well in all the stress bins for all the four cases.

## 4.2 Dependence of the triggering ratio, R, on Coulomb stress change

In Figure 2a, we show the variation of ratios $R^{k+}$ and $R^{k-}$ as a function of Coulomb stress change for PREF and RAND. For both PREF and RAND, $R^{k+}$ and $R^{k-}$ increase with increasing $|\Delta CFS_{ij}|_k^{median}$. Moreover, $R^{k+}$ seems to be larger than the corresponding $R^{k-}$ in both cases. We test the significance of this observation with a right tailed Ranksum test. In Figure 2b, we show the $p_{val}$ of the tests corresponding to PREF (red dots) and RAND (blue crosses). For the clarity of the figure, we have defined the lower limit for the $p_{val}$ to be $10^{-10}$. We have chosen two significance levels of 0.01 and 0.05 to demonstrate the stability of our results, which are shown as dotted and solid lines respectively. For both PREF and RAND, we find that $R^{k+}$ is significantly larger than the corresponding $R^{k-}$ above a stress threshold of ~10 Pa and ~24 Pa respectively, which does not seems to be affected by the choice of the significance level.

We also find that the value of $R^{k+}$ corresponding to PREF seems to be larger than in for RAND for all stress bins above ~9 Pa. On the other hand, $R^{k-}$ for PREF seems to be smaller than for RAND for all stress bins above ~6 Pa. We also verify these observations using the Ranksum test as described before (Figure S3).

Finally, we also note in Figure 2a that $R^{k+} - R^{k-}$ increases with increasing $|\Delta CFS_{ij}|_k^{median}$ in both cases of PREF and RAND. Moreover, $R^{k+} - R^{k-}$ corresponding to PREF is larger than for RAND.

## 4.3 Variation of Coulomb Indices (CI) as a function of Coulomb stress change and waiting time

Figure 3(a) and Figure 4(a) show the variation of CI values as a function of waiting time and Coulomb stress change for PREF and RAND, respectively.

To check for the statistical significance of the CI values in both cases over the respective MFCI value given the uncertainty in estimating these two quantities, we perform the statistical test proposed in Section 3.3.5. Figure 3b and Figure 4b show the map of $p_{val}$ obtained from this test as a function of Coulomb stress change and waiting time for PREF and RAND respectively. As before, we have artificially set the lower-limit of the $p_{val}$ obtained from both tests to $10^{-10}$.

Both CI value and significance map for both PREF (Figures 3a-b) and RAND (Figures 4a-b) show the existence of three Coulomb stress regimes. The first regime corresponds to the smallest Coulomb stress changes smaller than ~6 Pa and ~8 Pa for PREF and RAND respectively. In this stress regime, the CI values for both cases are always smaller or indistinguishable from their respective Mean-Field CI values. In the second regime, Coulomb stress change larger than ~6 Pa and waiting time smaller than ~158 days for PREF and Coulomb stress change larger than ~8 Pa and waiting time smaller than ~108 days for RAND, the CI values for both cases are significantly larger than respective Mean-Field CI of 0.54 and 0.52 respectively. In the third regime, for waiting times larger than ~158 days for PREF and ~108 days for RAND, the CI values seems to be indistinguishable from the Mean-Field CI values. All the three regimes have been annotated both in Figures 3b and 4b using double-headed arrows.

In Figure 5, we show the variation of the median $CI_k(t=0)$ as a function of $|\Delta CFS_{ij}|_k^{median}$ for both PREF (red circle) and RAND (blue star). We observe that $CI_k(t=0)$ increases with increasing $|\Delta CFS_{ij}|_k^{median}$. Moreover, $CI_k(t=0)$ corresponding to PREF is systematically larger than in the case of RAND.

We also find that the CI values corresponding to PREF are almost always larger than the CI values corresponding to RAND for $|\Delta CFS_{ij}|_k^{median}$ larger than ~11 Pa with some exceptions at large waiting times, where the fitting model shows less adequacy.

## 5. Discussion

### 5.1 Evidences for and against static stress triggering

#### 5.1.1 Findings supporting static stress triggering

For PREF, significantly larger values of $R^{k+}$ compared to $R^{k-}$ (Figure 2a and 2b) point towards preferential triggering in areas that received a positive Coulomb stress change from preceding earthquakes, compared to areas that were supposedly relaxed by negative Coulomb stress changes. Moreover, the increase of $R^{k+} - R^{k-}$ and $R^{k+}$ with increasing Coulomb stress change is consistent with static triggering.

Last but not the least, significantly larger CI values compared to MFCI for Coulomb stress change larger than ~6 Pa and waiting time smaller than ~158 days for PREF and Coulomb stress change larger than ~8 Pa and waiting time smaller than ~108 days for RAND (Figures 3-4) correlate with preferential triggering in positive stress bins for waiting times smaller than few hundred days, which cannot be explained by the time independent geometry of the fault network.

#### 5.1.2 Findings in apparent contradiction with static stress triggering

We also find evidence for Omori decay in the negative stress bins (Figure 1). The Omori decay is thought to be a signature of triggering and should therefore not be observed for negative stress bins at all.

Further, we find that $R^{k-}$ increases with the amplitude of Coulomb stress change (Figure 2a). This observation implies that larger negative Coulomb stress changes lead to larger triggered to background rate ratio. This observation again is in contradiction with the static triggering hypothesis.

However, the above observations can be accounted for by remarking that there exist uncertainties in the sign of the computed Coulomb stress changes, which leads to mixing of waiting times between a pair of positive and negative Coulomb stress bins with similar amplitude of stress change. While the uncertainty in the sign of the computed Coulomb stress change exists at all distances due to the focal plane ambiguity, or the uncertainty in the location events and in the orientations of the failure plane and slip vector, it is further aggravated in regions close to the source, which correspond to higher Coulomb stress changes. It then depends severely on unknown details of the slip distribution, compared to regions far away from the source. This unavoidable mixing predicts, in agreement with observations, that similar trends should be observed for the parameters inverted from the waiting time distribution in both cases.

Alternatively, the apparent contradiction of observations with the static triggering hypothesis could be accounted for if we accept that static triggering is not the sole triggering mechanism and works in conjunction with other triggering mechanisms such as dynamic triggering, which has often been invoked in literature to explain the apparent absence of stress shadows [Felzer et al., 2006; Felzer and Brodsky, 2005]. A possible qualitative model could be that the passage of seismic waves initiates a relatively isotropic triggering around the source according to the Omori law that is further modulated by the anisotropic static stress field. Furthermore, as the distance between the source and the targets increase, the static stress change differences between positive and negative stress bins diminish. As a result, the ability of static stress changes to modulate the triggered seismicity initiated by the passage of seismic waves would also diminish with decreasing amplitude of the stress changes, which could explain the decrease in differences between $R^{k+}$ and $R^{k-}$ with decreasing amplitude of the Coulomb stress change.

Finally, the aforementioned contradictions could also be explained solely in the framework of static triggering if we consider that the sources can not only trigger the targets directly but also in multiple steps [Saichev et al., 2005]: a primary source triggers a secondary source that triggers the target. Note that any number of intermediate steps could exist between the primary source and the target. It is then possible that even though the primary source and the target are directly connected via a negative Coulomb stress change, the actual pathway of triggering followed a positive Coulomb stress change at each step. Since the whole pathway is not accounted for in our analysis, this would then give rise to observations that would appear contradictory to the static triggering hypothesis.

While all the aforementioned scenarios solely, or in combination, explain the observations of similar behaviors for both positive and negative Coulomb stress bins, they would lead to inconclusive results if there were no other properties allowing us to distinguish the positive and negative stress bins. In fact, we observe much larger effects of the Coulomb stress change in positive stress bins than in negative stress bins. Indeed, this is precisely what we should expect if (i) the static stress triggering mechanism is present and (ii) one or all of the abovementioned scenarios were true.

## 5.2 Comparison of results obtained for PREF and RAND

We have larger mixing between positive and negative stress bins for RAND than for PREF, as we randomly chose the nodal planes of all earthquakes for RAND. Larger mixing of waiting times between positive and negative stress bins in the RAND case can explain the observations that $R^{k+}$ for PREF are larger than for RAND, while $R^{k-}$ for PREF are smaller than for RAND (Figure 2a and Figure S3). The mixing effect neutralizes the differences in seismicity rate in positive and negative stress bins. As a result, we find that the difference between $R^{k+}$ and $R^{k-}$ is larger in the PREF case than in the RAND case. This mixing effect also leads to significantly larger CI values in the PREF case than in the RAND case.

The comparison of the results between these two cases also outlines the importance of the choice of nodal planes before computing Coulomb stress changes. Thus, correctly choosing a nodal plane would increase our ability to observe the effects of sign of static stress changes. Moreover, this comparison allows us to quantitatively verify the predictions of the argument of unavoidable mixing between positive and negative stress bins offered in section 5.1.2 to explain the evidences apparently contradictory to the static triggering hypothesis.

## 5.3 Minimum Coulomb stress change threshold for static triggering

In Table 5, we list the several Coulomb stress change thresholds observed for the different seismicity parameters we measured (see also Figures 2-4). We find that Coulomb stress thresholds observed for different metrics agree well with each other. Note that all the observed stress change thresholds above which we detect the influence of the sign of Coulomb stress change are significantly lower than those previously published [Reasenberg and Simpson, 1992; Hardebeck et al., 1998; Toda et al., 1998; Anderson and Johnson, 1999]. However, we find that even below ~10 Pa for PREF and ~24 Pa for RAND, the log of the ratio of triggered to background seismicity rate is significantly larger than 0, the expected value in the case of no triggering, independently of the sign of Coulomb stress change. Combining these two evidences, we postulate that even though significant triggering is observed in all the stress bins, we only evidence a detectable modulating influence of the sign of static stress changes, a signature of static triggering, above a certain stress threshold because of the possible influence of uncertainties in Coulomb stress changes and the sensitivity of the metric used to detect that influence. Such an observation can also be rationalized if we consider, for instance, that static triggering might not be the sole triggering mechanism and works in conjunction with other triggering mechanisms, such as dynamic triggering (see section 5.1.2).

Note that some other studies [Ogata, 2005; Ziv and Rubin, 2000] have also shown evidences against a minimum threshold necessary for triggering. Specifically, Ziv and Rubin [2000] find significant triggering for Coulomb stress changes <1 kPa.

Finally, the stress thresholds for criteria 1-4 (Table 5) correspond roughly to a distance (in units of source length) of 5.7, 4.5, 6 and 6.5 respectively. Beyond these distances, we do not observe the influence of the sign of Coulomb stress changes. These distances can possibly define the size of the aftershock zone in which using the sign of the Coulomb stress changes can possibly improve the forecasting abilities of spatially isotropic models such as Epidemic Type Aftershock Sequence (ETAS) models.

**5.4 Aftershock duration**

The time $\tau$ is related to the average duration of aftershock sequences (Figure 6; also see section 3.3.1 and Table 1). It has been inverted directly from the waiting time distribution and can be defined as the average characteristic time beyond which the rate of aftershocks decays exponentially to the background seismicity rate. It can thus be interpreted as the largest time scale until which the memory of past stress changes survives, thus being an effective Maxwell time of the relaxation process [Sornette and Ouillon, 2005; Ouillon and Sornette, 2005].

Figure 6 shows that the characteristic times for triggered seismicity, $\tau^{k+}$ and $\tau^{k-}$, slightly increase with increasing absolute amplitude of the median Coulomb stress change $|\Delta CFS_{ij}|_k^{median}$, both for PREF and RAND. The increase is marginal, and is only about 0.25 decades over 5 decades of variation in stress change amplitude. The median value of the characteristic time varies between ~95 days and ~180 days.

Also note that the inverted characteristic times of the waiting time distribution agree with the onset time of Regime 3 (~158 days for PREF and ~108 days for RAND) evidenced in Figures 3 and 4. The CI values that are significantly larger than the MFCI prior to the onset time become indistinguishable from MFCI post this time. In other words, beyond the onset time of Regime 3, which agrees with $\tau$, no evidence of static triggering is observed. This further provides support for $\tau$ representing the average characteristic time scale until which the memory of past static stress changes survives.

Several mechanisms could be proposed for the fading of the memory of past stress changes. First, in a Maxwell material suddenly subjected to a deformation step, the stress decays with a characteristic time $\frac{\eta}{E}$, where, $\eta$ is the coefficient of viscosity and E is the elastic Young modulus. This characteristic time is a material property and does not depend on the amplitude of the stress change. Plugging in the values of the characteristic time and the elastic modulus, E = 32 GPa, we find the viscosity to vary between ~ $2.7 \times 10^{17}\ Pa\ s$ and ~ $4.7 \times 10^{17}\ Pa\ s$. This low value seems to be inconsistent with the very high viscosity estimates for the crust of the order of ~$10^{24}$ Pa s at seismogenic depths [Bills et al., 1994].

Overprinting of stress changes caused by following earthquakes could also erase the memory of the past stress. In that case, larger stress change areas should correspond to larger $\tau$ compared to areas with smaller Coulomb stress change threshold. However, we also find that areas of larger Coulomb stress changes are associated with higher seismicity rates (Figure 2a). As a result, areas with high Coulomb stress changes would receive more imprinting stresses, which would facilitate the deletion of the memory of past stress changes. The interplay of these two effects could lead to an approximately constant characteristic time $\tau$.

Given the range of $\tau$ between ~95 days and ~180 days, the observations of long aftershock sequences of several large earthquakes (e.g. Stein and Liu, 2009) might seem paradoxical. This paradox can be explained if we consider that $\tau$ is just an average parameter. In reality, it might feature spatio-temporal variation which can lead to exceptionally large or small characteristic times in some specific regions due to influence of local geological processes. Moreover, we should also consider that the studies that have claimed to observe exceptionally long aftershock sequences (longer than the usual length of instrumental catalogs) [Ebel et al., 2000; Stein et al., 2009] feature huge uncertainties on the estimated background seismicity rate prior to the main shock, a key parameter necessary for estimating the aftershock duration and occur in different tectonic settings. Last but not the least, Figure 2 from Davidsen et al. [2015] shows that both "bare" and "dressed" aftershock sequences corresponding to both Landers and Hector Mine earthquakes clearly display the signature of an exponentially tapered Omori law with a characteristic time of the order of a few hundred days. Also, in Text S3, we further show that $\tau$ estimated by the means of a modified Epidemic Type Aftershock Sequence (ETAS) model agrees well with the one inverted from the waiting time distributions.

The knowledge of $\tau$ sets an average, non-arbitrary time boundary for case studies that are based on Coulomb stress changes caused by main shocks. Most importantly, our results denote that forecasts based on static stress changes will have poor predictive skill beyond times that are much larger than a few hundred days on average.

## 6. Summary and Conclusions

We conducted robust tests of the static triggering hypothesis by considering Coulomb stress changes by all events with magnitude larger than 2.5 recorded in the state-of-the-art focal mechanism catalog of southern California. We also considered the often-ignored uncertainties in the Coulomb stress changes due to those in location and focal mechanism of the earthquakes in the available catalog.

We performed a two-fold test of the static triggering hypothesis. In the first case, we resolved the focal plane ambiguity of the earthquakes, a problem often disregarded in previous Coulomb stress change studies, by first reconstructing the fault planes within the predefined clusters present in the relocated catalog of southern California using a pattern recognition method inspired by Ouillon and Sornette [2011], combined with a condition of slip consistency within individual fault segments. In the second case, we chose the nodal planes randomly between the two available choices.

We modeled the waiting time distribution between earthquakes conditioned on the sign and amplitude of the Coulomb stress change both, for negative and positive cases, using the exponentially tapered Omori law. We used the parameters of the waiting time distribution to evaluate different quantities relevant to testing the static triggering hypothesis.

We found that compelling evidence exists for the static triggering hypothesis above consistently observed small stress thresholds (about 10Pa). We conclude that by resolving the focal plane ambiguity, we are able to see the signature of static triggering more clearly, which indicates the importance of an informed choice of fault planes. The evidence in favor of the static triggering hypothesis is further reinforced by our finding that the influence of the sign of static stress changes is independent of the values of the coefficient of friction, Skempton's coefficient and magnitude threshold (Text S2). We find very weak correlations between the Coulomb Index and the coefficient of friction, which indicate that the role of normal stress in triggering is rather limited.

Last but not the least, we find the characteristic time of the stress change memory of a single event to be nearly independent of the amplitude of the Coulomb stress change and to vary within the range from ~95 to ~180 days. It sets an average non-arbitrary period for case studies that are based on Coulomb stress changes caused by main shocks. Most importantly, our results indicate that forecasts based on static stress changes will have poor predictive skill beyond times that are larger than a few hundred days on an average.

**Acknowledgement:**

The three dataset used in this study can be obtained from the websites: http://scedc.caltech.edu/research-tools/alt-2011-yang-hauksson-shearer.html, http://scedc.caltech.edu/research-tools/alt-2011-dd-hauksson-yang-shearer.html and http://equake-rc.info/.

S.N. would like to acknowledge Men Andrin Meier for introducing him to the topic and providing him with his codes. S.N. also benefitted from intriguing discussions with M. J. Werner and countless valuable suggestions from Yavor Kamer. We would also like to thank the Editor, the Associate Editor and the 3 anonymous reviewers for their critical comments that greatly helped to improve our manuscript.

**Table 1**: List of symbols and acronyms frequently used in the main text in their order of appearance.

| Symbol/Acronym | Definition |
|---|---|
| YANG | Focal Mechanism catalog of Southern California [Yang et al., 2012] |
| HAUK | Relocated catalog of Southern California [Hauksson et al., 2012] |
| $m_t$; $b_{val}$ | Magnitude threshold above which Gutenberg-Richter law [Gutenberg and Richter, 1954] holds; Measure of relative frequency of small to large earthquake. |
| $\mu, B, \mu_B$ | Coefficient of Friction (0.6); Skempton's Coefficient (0.75); Shear Modulus (32 GPa) |
| $\Delta CFS_{ij}$; $\|\Delta CFS_{ij}\|_k^{median}$ | Coulomb stress change caused by i$^{th}$ source at the location of j$^{th}$ target; Median of $\|\Delta CFS_{ij}\|$ in the k$^{th}$ stress bin, where k varies between 1 and n$_{bin}$(=50) |
| $t_{ij}^k$; $t_{ij}^{k+}$ & $t_{ij}^{k-}$ | Waiting time between all source target pairs present in k$^{th}$ stress bin; waiting time between source target pairs present in kth stress bin given the associated Coulomb stress change is positive and negative respectively. |
| $N^k$; $N^{k+}$ & $N^{k-}$ | Total Number of source-target pairs in the k$^{th}$ stress bin; Number of source target pairs for which associated Coulomb stress change is positive and negative respectively. Also, $N^k = N^{k+} + N^{k-}$ |
| W | Probability density function of waiting time (Equation 3). |
| $\lambda$; $\lambda^{k+}$ & $\lambda^{k-}$ | Parameters of the waiting time distribution: {K, c, $\rho$, $\tau$, B, $\theta$}; Parameters of the conditional waiting time distribution in the k$^{th}$ stress bin given that the associated Coulomb stress change is positive and negative respectively. |
| $\tau$; $\tau^{k+}$ & $\tau^{k-}$ | Characteristic time for the loss of the stress change memory of a single event; Characteristic time for the k$^{th}$ stress bin given that the associated Coulomb stress change is positive and negative respectively. |
| $MLL$; $MLL^{k+}$ & $MLL^{k-}$ | Maximum Log Likelihood per data point; Maximum Log Likelihood per data point in the k$^{th}$ stress bin given that the associated Coulomb stress change is positive and negative respectively. |
| $R$; $R^{k+}$ & $R^{k-}$ | Ratio of instantaneous triggered to background rate; Ratio of instantaneous triggered to background rate given that the |
| | associated Coulomb stress change is positive and negative respectively. |
| $CI_k(t)$; MFCI | Fraction of causal pairs in the earthquake catalog with positive Coulomb stress change interaction in the $k^{th}$ stress bin at time $t$ (Equation 6); Fraction of positive Coulomb stress change interactions in earthquake catalog that has been randomly shuffled in time. |
| $\alpha$; $p_{val}$ | Significance level chosen to test a null hypothesis; Probability of obtaining the test statistics at least as extreme as that was actually observed assuming that the null hypothesis is true. |
| PREF; RAND | Case corresponding to preferred choice of nodal planes according to method proposed in section 3.2.2; Case corresponding to random choice of nodal planes. |

**Table 2**: List of earthquakes for which a source model is available.

| Name of Earthquake | References |
|---|---|
| Elmore Ranch | Larsen et al. [1992] |
| Hector Mine | Salichon et al. [2004] |
| | Kaverina et al. [2002] |
| | Jonsson et al. [2002] |
| Joshua Tree | Hough and Dreger [1995] |
| | Bennet et al. [1995] |
| Landers | Zeng and Anderson [2000] |
| | Wald and Heaton [1994] |
| | Hernandez et al. [1999] |
| | Cotton and Campillo [1995] |
| | Cohee and Beroza [1994] |
| Big Bear | Jones and Hough [1995] |
| North Palm Spring | Mendoza and Hartzell [1988] |
| | Hartzell [1989] |
| North Ridge | Dreger [1994] |
| | Hartzell et al. [1996] |
| | Hudnut et al. [1996] |
| | Shen et al. [1996] |
| | Wald et al. [1996] |
| | Zeng and Anderson [2000] |
| Sierra Madre | Wald [1992] |
| Superstition Hill | Wald et al. [1990] |

|  | Larsen et al. [1992] |  |
|---|---|---|
| Whittier Narrows | Hartzell and Iida [1990] |  |

**Table 3:** Empirical relations from Wells and Coppersmith [1994] used to estimate the subsurface-rupture length (L), rupture width (W) and amplitude of average slip ($\bar{u}$) given the magnitude, M, and faulting style of any earthquake; All the relations have the general form $10^{a*M+b}$.

|  |  | a | b |
|---|---|---|---|
| **Rupture Length (L)** | **Strike Slip** | 0.62 | -2.57 |
|  | **Reverse** | 0.58 | -2.42 |
|  | **Normal** | 0.50 | -1.88 |
|  | **Oblique** | 0.59 | -2.44 |
| **Rupture Width (W)** | **Strike Slip** | 0.27 | -0.76 |
|  | **Reverse** | 0.41 | -1.61 |
|  | **Normal** | 0.35 | -1.14 |
|  | **Oblique** | 0.32 | -1.01 |
| **Average Displacement ($\bar{u}$)** | **Strike Slip** | 0.9 | -6.32 |
|  | **Reverse** | 0.08 | -0.74 |
|  | **Normal** | 0.63 | -4.45 |
|  | **Oblique** | 0.69 | -4.80 |

**Table 4:** List of optimal parameters of the best fit of the waiting time distributions for different cases (Figure 1).

| Case | K | $\rho$ | c (days) | $\tau$ (days) | B | $\theta$ (days) |
|---|---|---|---|---|---|---|
| **PREF, $+\Delta CFS$** | 0.40 | 0.81 | 0.039 | 133 | $2.2\times10^{-4}$ | 3011 |
| **PREF, $-\Delta CFS$** | 0.13 | 0.72 | 0.049 | 131 | $2.2\times10^{-4}$ | 2876 |
| **RAND, $+\Delta CFS$** | 0.24 | 0.78 | 0.042 | 137 | $2.3\times10^{-4}$ | 2924 |
| **RAND, $-\Delta CFS$** | 0.22 | 0.75 | 0.049 | 133 | $2.4\times10^{-4}$ | 3153 |

**Table 5:** Coulomb stress change thresholds observed for different seismicity parameters measured from waiting time distributions for 50 Coulomb stress bins; Criteria (1-2) correspond to the ratio of triggered to background seismicity rate (R) whose variation is shown in Figure 2a-b; Criteria (3-4) correspond to the fraction of positive Coulomb stress interaction (CI) whose variation and significance over the mean-field CI value is shown in Figure 3-4.

| Index | Criteria | Coulomb stress threshold |
|---|---|---|
| 1 | $R_{PREF}^{k+} > R_{PREF}^{k-}$ | ~10 Pa |
| 2 | $R_{RAND}^{k+} > R_{RAND}^{k-}$ | ~24 Pa |
| 3 | $CI_{PREF} > MFCI_{PREF}$ | $\Delta CFS > 6\ Pa;\ t_{ij} < \sim 158\ days$ |
| 4 | $CI_{RAND} > MFCI_{RAND}$ | $\Delta CFS > 8\ Pa;\ t_{ij} < \sim 108\ days$ |

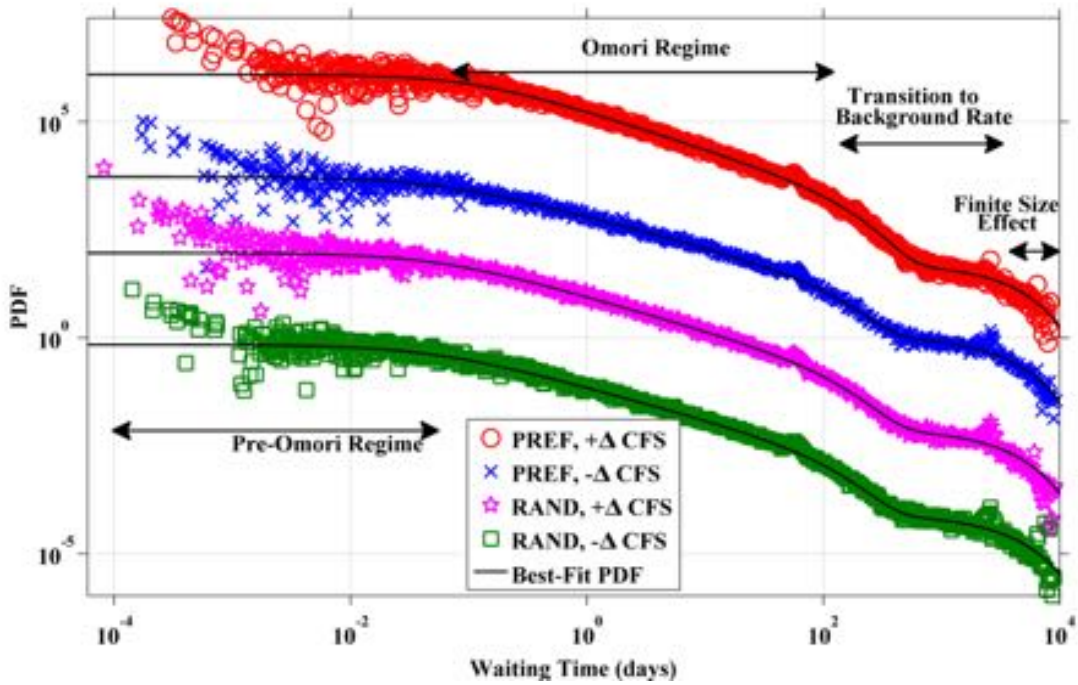

**Figure 1**: Empirical and proposed best fit waiting time distributions for positive and negative stress bins and for the preferred and random focal mechanism choice (PREF and RAND) respectively. Scattered markers represent the empirical distributions obtained using a kernel density estimation. Smooth lines show the best fitted waiting time distribution obtained by maximizing the log likelihood with corresponding optimal parameters shown in Table 4. Both, the empirical and the fitted waiting time distribution, have been normalized to 1 and then translated along Y-axes by a factor of 100 for comparative visualization. The median Coulomb stress change in all cases is ~42,000 Pa.

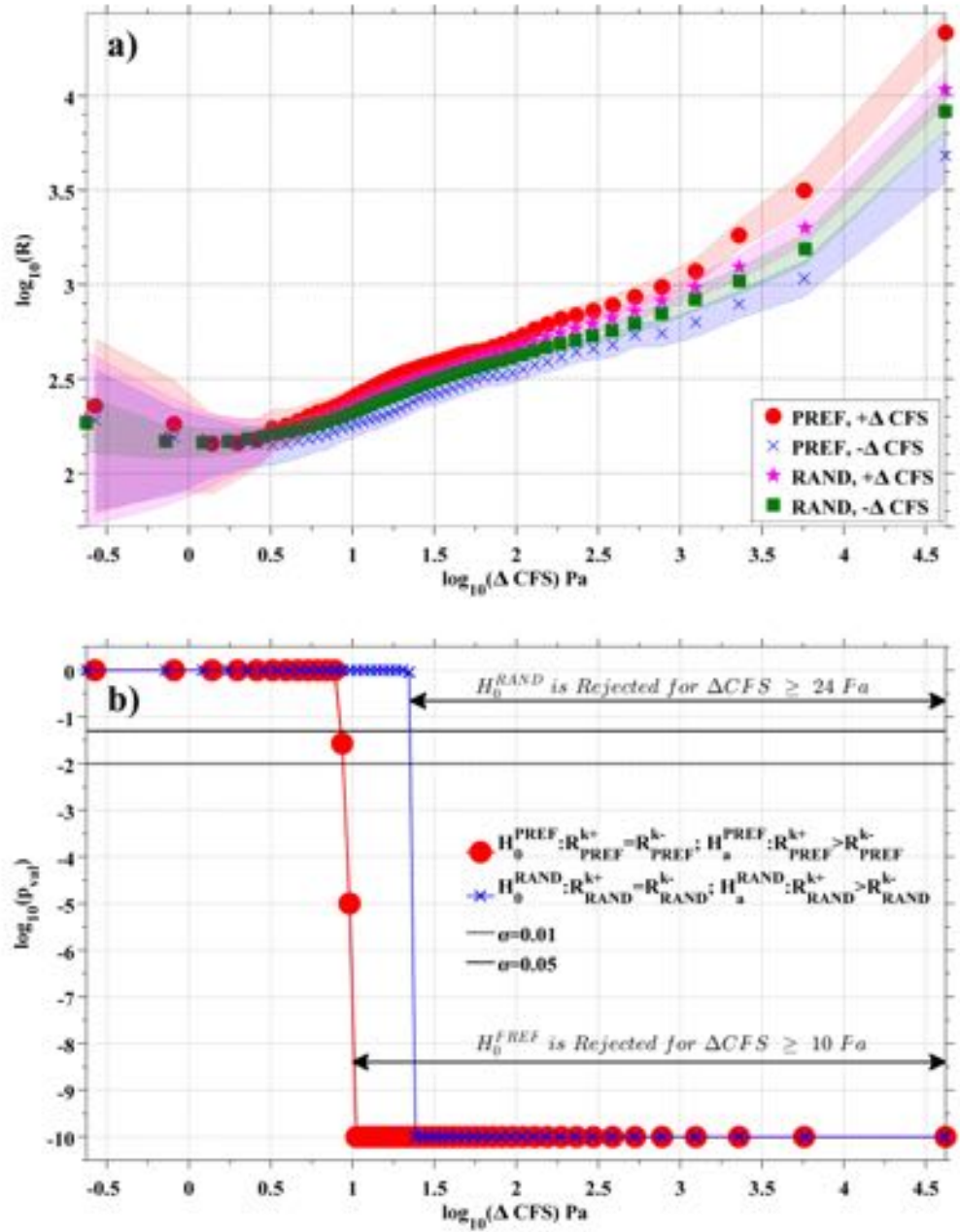

**Figure 2**: **(a)** $R^{k+}$ and $R^{k-}$ (Ratio of instantaneous triggering to background rate) vs. Coulomb stress change for both PREF and RAND; the legend shows the labels corresponding to the four cases; the median value corresponding to each case is plotted using markers; shaded region plotted in the same colors as the markers show the 95% confidence interval delineated by the 2.5% and 97.5% quantiles; **(b)** $log_{10}(p_{val})$ of Ranksum test vs. Coulomb stress change for PREF (Red circles) and RAND (Blue crosses); The null hypothesis and the alternative one in both cases is shown in the legend; the significance level of 0.01 and 0.05 is shown in grey line and black line respectively; The Coulomb stress change threshold above which the null hypothesis is rejected is 10 Pa and 24 Pa for PREF and RAND respectively.

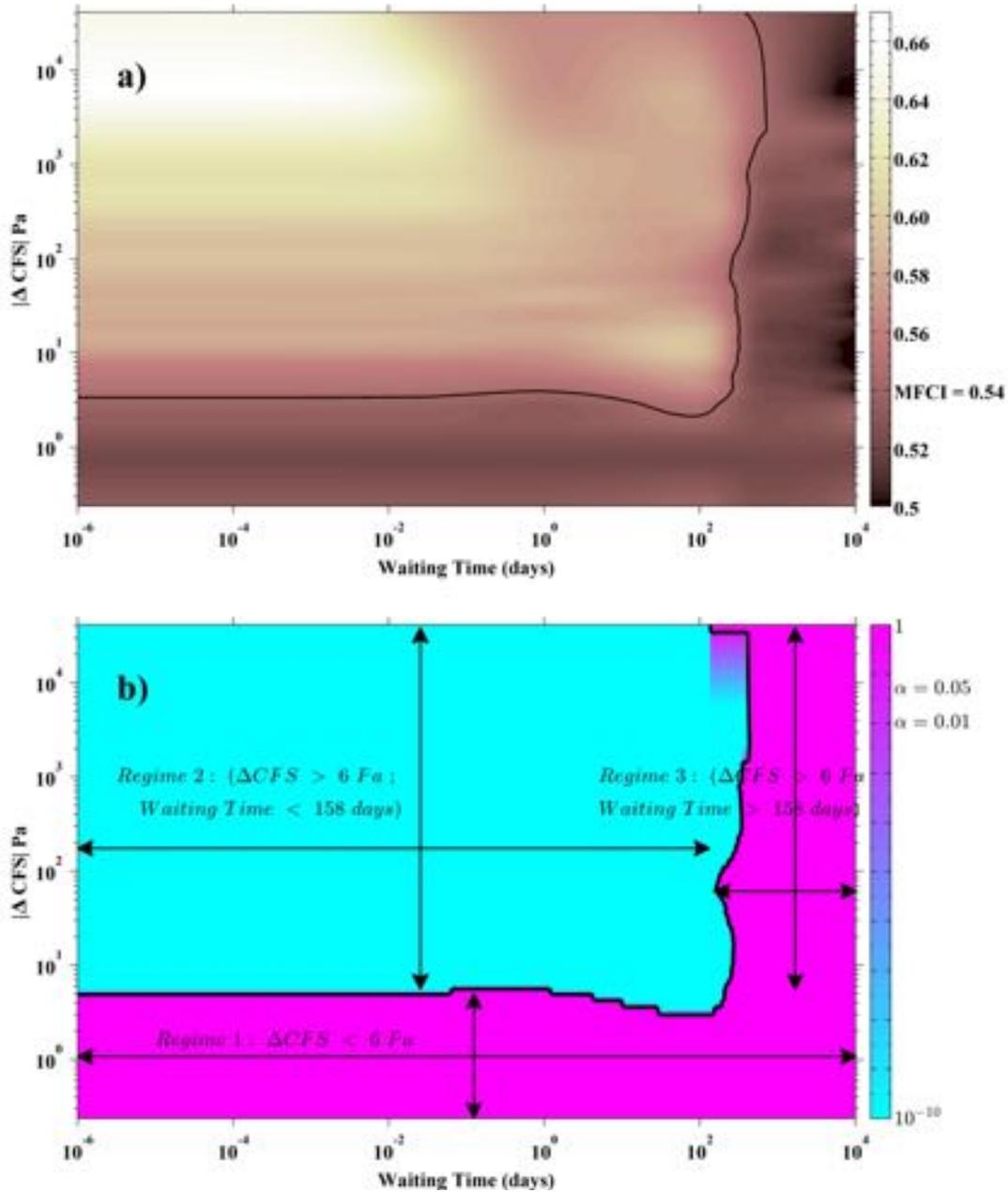

**Figure 3**: **(a)** Map of observed CI value as a function of waiting time and Coulomb stress change for PREF; The color corresponding to the Mean Field CI value (MFCI), which equals 0.54, is shown on the color bar; The contour line corresponding to the MFCI is shown as the black solid line on the map; **(b)** Map of $p_{val}$ of the Ranksum test as a function of waiting time and Coulomb stress change showing the significance of CI values for PREF over the corresponding MFCI; The colors corresponding to the significance level 0.01 and 0.05 are shown in the color bars; The black contour line on the map mark the $p_{val}$ equal to the significance level $\alpha = 0.01$; Double headed arrows mark the three regimes that are apparent on the map.

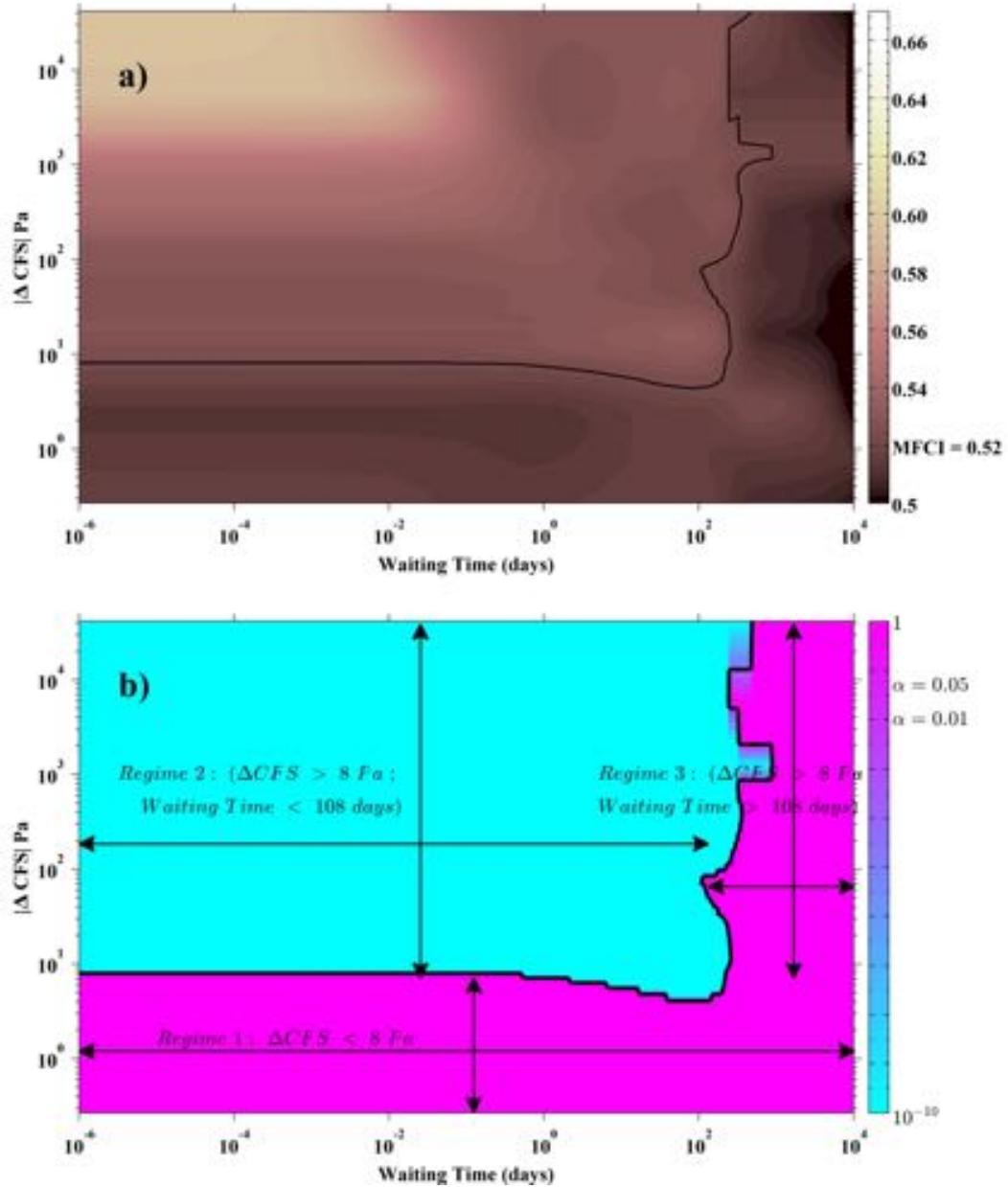

**Figure 4**: **(a)** Map of observed CI value as a function of waiting time and Coulomb stress change for RAND; The color corresponding to the Mean Field CI value (MFCI), which equals 0.52, is shown on the color bar; The contour line corresponding to the MFCI is shown as the black solid line; **(b)** Map of $p_{val}$ of the Ranksum test as a function of waiting time and Coulomb stress change showing the significance of CI values for RAND over the corresponding MFCI; The colors corresponding to the significance level 0.01 and 0.05 are shown in the color bars; The black contour line on the map mark the $p_{val}$ equal to the significance level $\alpha = 0.01$; Double headed arrows mark the three regimes that are apparent on the map.

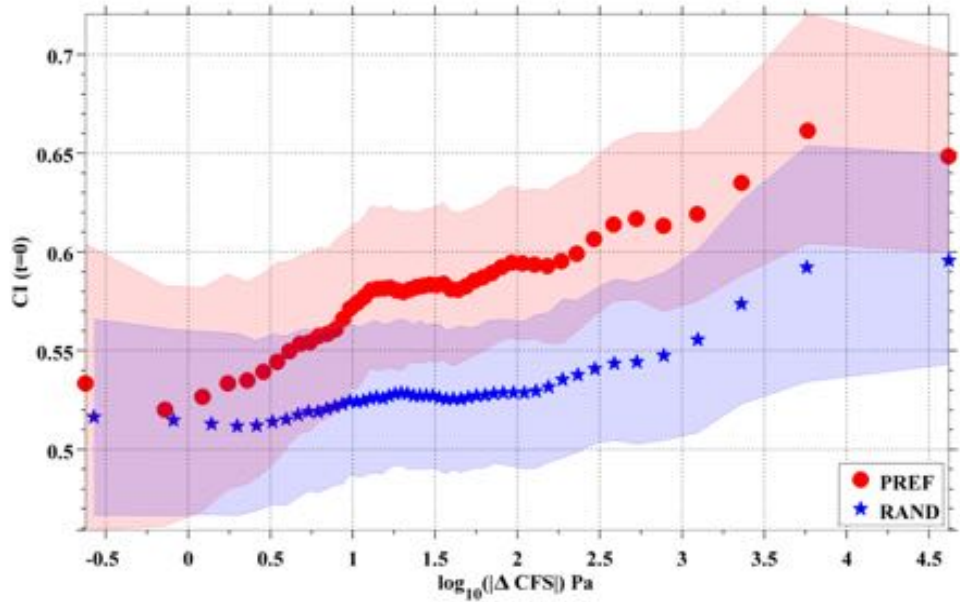

**Figure 5:** CI(t=0) as function of Coulomb Stress change for both preferred and random choice (PREF and RAND, see legend), The median value corresponding to each case is plotted using markers. The shaded regions plotted in the same colors as the markers show the 95% confidence intervals delineated by the 2.5% and 97.5% quantiles.

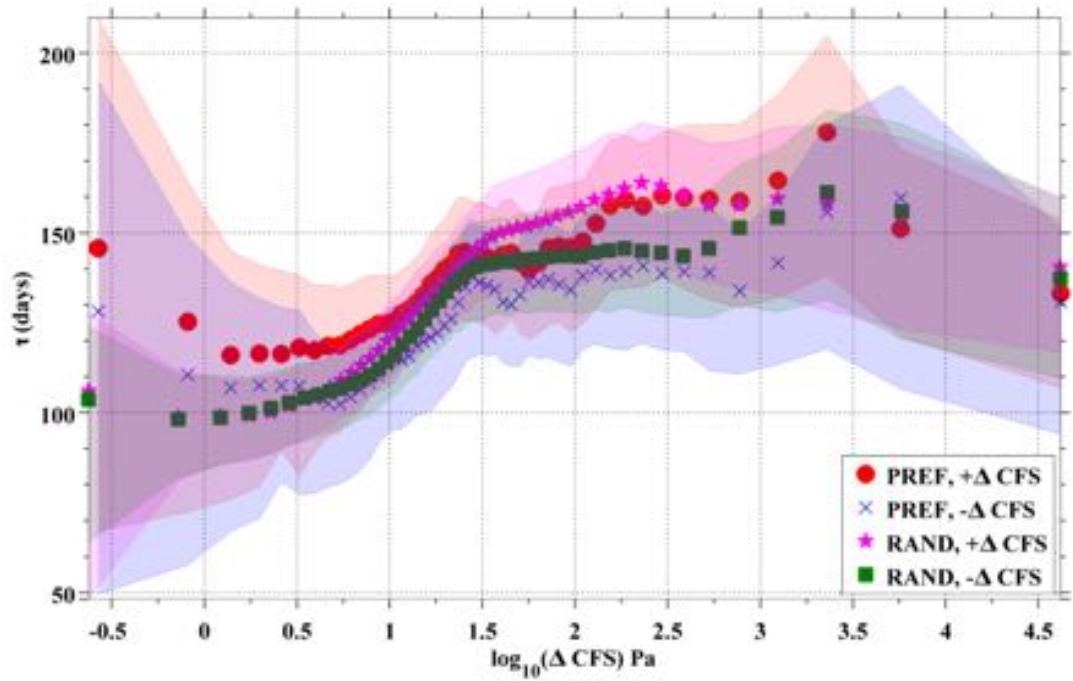

**Figure 6**: $\tau^{k+}$ and $\tau^{k-}$ vs. Coulomb stress change for both PREF and RAND. The median value corresponding to each case is plotted using markers. The shaded region shown in the same colors as the markers depict the 95% confidence intervals.

# Supporting Information for

**Systematic Assessment of the Static Stress-Triggering Hypothesis using Inter-earthquake Time Statistics**

Shyam Nandan[1], Guy Ouillon[2], Jochen Woessner[3,1], Didier Sornette[4] and Stefan Wiemer[1]

[1]ETH Zürich, Swiss Seismological Service, Sonneggstrasse 5, 8092 Zürich, Switzerland

[2]Lithophyse, 4 rue de l'Ancien Sénat, 06300 Nice, France

[3]Risk Management Solutions Inc., Stampfenbachstrasse 85, 8006 Zürich, Switzerland

[4]ETH Zürich, Department of Management, Technology and Economics, Scheuchzerstrasse 7, 8092 Zürich, Switzerland

**Contents of this file**

Text S1 to S4

Figures S1 to S13

Table S1

References

**Text S1: Application of Ouillon and Sornette [2011] for optimal Gaussian partitioning**

As described in section 3.2, each cluster of the HAUK-catalog needs to be further partitioned into sub-clusters of earthquakes that could be designated as occurring on a single fault segment (see Figure S4). The general strategy for achieving this is the following. For a given number of sub-clusters, we use a Gaussian mixture approach implemented through an expectation-maximization (EM) procedure to obtain the optimal decomposition of a given cluster into well-defined individual Gaussian distributions. We repeat the same procedure progressively, increasing the number of sub-clusters. The optimal number of sub-clusters corresponds to the configuration with minimum Bayesian Information Criterion (BIC) [Bishop, 2006].

**Text S1.1: Prior Identification of Outliers**

As observed by Ouillon and Sornette [2011], several events remain isolated and possibly do not belong to any cluster (Figure S4). Even though each of them must belong to an underlying fault, the existence of those faults is be revealed by seismicity due to the limited number of earthquakes associated to them. Including these isolated events in the clustering process would tend to identify sub-clusters with three or less earthquakes, leading to near-singular log likelihoods, preventing us from computing correctly the BIC.

Ouillon and Sornette [2011] proposed to remove this uncorrelated seismicity using a filtering method, hereafter labeled as F1, which compares the distribution of volumes of tetrahedra ($V_{tetra}$) (defined by event and its three nearest neighbors) in the natural catalog to a synthetically generated reference distribution for uncorrelated seismicity (Section 8.2 of Ouillon and Sornette, 2011).

We introduce some minor modifications to the original approach of Ouillon and Sornette [2011] to define a modified filtering method, labeled as F2. First of all, we consider the convex hull of the observed cluster, and generate within that hull the same number of uncorrelated (randomly distributed) events. The convex hull is defined as the minimum convex set that contains the earthquakes and naturally defines a unique boundary for the given set of points. We then find the volume of the tetrahedron formed by each point and its three nearest neighbors in the synthetic and in the natural catalog.

In order to improve on Ouillon and Sornette [2011], we also compute the maximum distance ($d_{max}$) between each earthquake and its three nearest neighbors in both catalogs, as an event may be an outlier even if its associated tetrahedron possesses a very small volume. The latter situation occurs if the considered event is spatially isolated, yet its three closest neighbours are tightly clustered. In such a case, the event would not be labeled as an outlier and would participate in the clustering process, unavoidably leading to singularities in the log likelihood and eventually to sub-optimal spurious solutions. By also comparing the distribution of $d_{max}$ in the natural and synthetic catalogs, one can easily detect and remove such outliers.

Finally, all the earthquakes in the natural set are then qualified as uncorrelated events or outliers if one of the above two quantities exceeds the 95%ile confidence boundary of the respective quantities ($V_{tetra}$ and $d_{max}$) for the reference distribution (Figure S5a). On the other hand, Ouillon and Sornette [2011] define an event to be an outlier if the corresponding $V_{tetra}$ computed from the natural catalog exceeds the 5%ile confidence boundary of the reference distribution of $V_{tetra}$ alone (Figure S5b). Note that, both methods rely on the unavoidable arbitrary choice of confidence boundary for the definition of outliers.

**Text S1.2: Optimal Gaussian Partitioning of cluster #50106**

Figure S6 shows the variation of the BIC corresponding to the optimal partitioning of the selected set of events for an increasing number of sub-clusters. We have indicated with a red star the optimal number of sub-clusters for which the minimum BIC is achieved. We find that after accounting for the increasing complexity of the model with an increasing number of Gaussian components, the best partitioning of the earthquakes in cluster #50106 could be achieved using 10 Gaussian sub-clusters.

Figure S7 shows the horizontal cross-section (Depth=0) of the optimal configuration of the 10 Gaussian sub-clusters. We represent each sub-cluster using colored ellipsoids with three semi-axes equal to the $\sqrt{3 \times \gamma_i^2}$, where $\gamma_i^2$ ($i = 1, 2\ or\ 3$) represents the three eigenvalues of the corresponding covariance matrix. According to Ouillon et al. [2008], twice the value of the largest two semi-axes represent approximately the fault length and width respectively. Earthquakes assigned to each sub-cluster are shown using different marker types and colors (same as the parent ellipsoid). Note that, an earthquake is assigned to the Gaussian sub-cluster that attributes highest probability to the earthquake. For convenience, we have also labeled each fault with a different index. Figure S8a (East=0) and S8b (North=0) show the two vertical cross-sections of the same ellipsoids shown in Figure S7. The eigenvector of the Gaussian kernel corresponding to the smallest eigenvalue is defined as the normal vector ($N_{clust}$) of the underlying fault plane and fully encodes its orientation.

For any given number of sub-clusters, we perform the clustering analysis 3,000 times, starting from different initial configurations, and only choose the final partition corresponding to the smallest BIC.

**Text S1.3: Resolving the focal plane ambiguity**

STEP 1: Within each sub-cluster, we first consider the nodal plane of each earthquake's focal mechanism such that its normal vector is the closest to the normal vector to the sub-cluster, $N_{clust}$. This automatically provides a choice for the slip vector associated to each event. This step allows us to compute an initial guess for the average slip vector ($s^{avg}_{step1}$) on the reconstructed fault plane of the corresponding cluster. $s^{avg}_{step1}$ is computed by projecting the mean of the individual unit slip vectors.

STEP 2: We then compute, for each event, the deviation between $s^{avg}_{step1}$ and both unit slip vectors provided by its associated focal mechanism. The slip vector yielding the smallest deviation is then definitely considered as the slip vector associated to that event (which automatically determines the failure plane). This is done to reduce the influence of uncertainties on STEP 1, which sometimes provides a few events in a cluster slipping in an opposite direction to all the others. We then update the estimate of the average slip vector on the reconstructed fault plane to $s^{avg}_{step2}$, defined as the projection of the mean of unit slip vectors selected in the second step. The amplitude of $s^{avg}_{step2}$ thus ranges between 0 and 1. Amplitude closer to 1 (resp. to 0) implies that the preferred slip vectors of the earthquakes in the cluster are less (resp. more) scattered in their directions.

We further compare the consistency of the preferred slip vectors $s^{avg}_{step2}$ to the case where the failure plane for each earthquake in the cluster is chosen randomly. For this, we compute the projection of the mean of the preferred unit slip vectors on the reconstructed fault plane of each cluster, $s^{avg}_{random}$, when we randomly chose the preferred nodal plane for all earthquakes. We find a substantial improvement of the amplitude for $s^{avg}_{step2}$ compared to $s^{avg}_{random}$ as displayed by the histograms of amplitude for $s^{avg}_{step2}$ and $s^{avg}_{random}$, computed for all the sub-clusters of earthquakes (Figure S9, red and blue, respectively). The median value of the $s^{avg}_{step2}$ amplitude histogram is 0.85 while that of $s^{avg}_{rand}$ is 0.32. This implies that the preferred slip vectors selected using the two-steps approach are more tightly oriented than the preferred slip vectors selected using a random choice between the nodal planes.

The above procedure provides multiple choices for computing Coulomb stress changes. For each earthquake, we can use the normal and slip vectors corresponding to the individually selected nodal plane. We can also assign to each earthquake in a cluster a fixed unit normal vector, $N_{clust}$ and a fixed unit slip vector, $\frac{s^{avg}_{step2}}{|s^{avg}_{step2}|}$. The use of the second approach is justified if we consider that, in the presence of a locally homogenous stress field, a planar fault possesses a unique slip direction. As a result, all the earthquakes associated with this fault should display a unique focal mechanism. The variation in the focal mechanism of the earthquakes associated with this fault plane is then just the manifestation of uncertainties in the inversion process used to obtain them. In this paper, we have followed this second 'stacking' approach.

As a demonstration of how the method proposed above works we also show in Figure S7 the composite focal mechanisms associated with the 10 sub-clusters using colored beach balls. We have obtained each composite focal mechanism using $N_{clust}$ as the normal vector and $\frac{s^{avg}_{step2}}{|s^{avg}_{step2}|}$ as the slip vector. We also display the strike, dip and rake of the preferred nodal plane corresponding to each composite focal mechanism. The colors of the beach balls correspond to those of the reconstructed faults shown by the colored ellipsoids.

**Text S2: Sensitivity to the values of the friction coefficient, Skempton's coefficient and magnitude threshold**

All the evidences for static triggering presented in the main text correspond to fixed values of the friction coefficient ($\mu = 0.6$), Skempton's coefficient ($B = 0.75$) and magnitude threshold $m_t = 2.5$. Given the uncertainties in these quantities, we consider imperative to test the sensitivity of our results to their possible variations.

We consider both $\mu$ and B to vary within the range [0,1], which is based on generally reported values for these two quantities, while we allow $m_t$ to vary within [2,3].

Our general strategy is to fix all parameters to their originally considered values, except for the one whose effect we are trying to capture. We then vary the parameter under consideration within its predefined range. For instance, if we are assessing the effect of $\mu$, we fix B and $m_t$ to their original values of 0.75 and 2.5 respectively. We then vary $\mu$ within the range [0,1] with a step of 0.1. For each value of $\mu$, we then estimate the quantities $R, CI(t = 0)\ and\ MFCI$ one thousand times, by bootstrapping the locations and focal mechanisms uncertainties. Similarly, we compute these quantities as a function of B and $m_t$ with the other two parameters fixed.

In Figure S10a-S10b, we show the variation of the median values of $R, CI(t = 0)\ and\ MFCI$ for all possible cases as a function $\mu$. We find that the both R and $CI(t = 0)$ exhibit all the properties discussed in section 4.2 and 4.3, independently of the choice of $\mu$. The same holds for the other two parameters B **(Figure S11a-S11b)** and $m_t$ **(Figure**

S12a-S12b). These observations further reinforce the evidences in favor of the static triggering hypothesis. We also demonstrate that, independently of the choice of $\mu$, B and $m_t$, our ability to find supportive evidence for the static triggering hypothesis is stronger if we make informed choices of the fault planes. It is, however, important to note that we find significantly stronger evidences supporting static triggering hypothesis when we make informed choices of fault planes even for $\mu = 0$, the case in which the computed Coulomb stress changes should be independent of the choice of nodal planes for both source and receiver. But, the independence of Coulomb stress change of the choice of nodal planes only holds in case of point sources and is not valid for sources with finite dimensions, as in the case of present study. So, the significantly different results for PREF and RAND even for $\mu = 0$ presents no contradiction at all.

We also find from Figure S10b and Figure S11b that *CI(t=0)* seems to show a very weak negative correlation with respect to $\mu$ and no correlation at all with *B* for both PREF and RAND respectively. Note that other researchers have also evidenced such a weak correlation between Coulomb Index and $\mu$ [Catalli et al., 2013]. The existence of a weak correlation between Coulomb Index and $\mu$ and almost no correlation between Coulomb Index and $B$ implies that the effect of the effective normal stress in earthquake triggering is only moderate and the dominating role is played by the shear stress. Note that, Kagan and Jackson [1998] reached similar conclusions by comparing the spatial distribution of earthquakes prior and posterior to a mainshock under the premise that, for positive values of $\mu$, one would observe that aftershocks would concentrate in the direction of the P-axis rather than in the direction of the T-axis.

**Text S3: Estimating the characteristic time, $\tau$, using a modified Epidemic Type Aftershock Sequence (ETAS) model**

An appropriate way to measure $\tau$ could be to use a modified Epidemic Type Aftershock Sequence (ETAS) model [Ogata and Zhuang, 2006; Zhuang et al., 2004]. This model allows for inverting the parameters of "bare" kernels, which describe the way in which an earthquake directly triggers its aftershocks. We modify the "pure" Omori kernel, $\frac{1}{(t+c)^p}$, with an exponentially tapered Omori kernel, $\frac{e^{-t/\tau}}{(t+c)^p}$, and then perform the usual inversion of its parameters [Veen and Schoenberg, 2008].

We perform the inversion of the ETAS parameters for both catalogs (HAUK and YANG) considered in this study with $M \geq 2.5$, magnitude above which both catalogs are thought to be complete (Figure S1). We replace the usual space-time-magnitude triggering kernel in the ETAS model [Ogata and Zhuang, 2006; Zhuang et al., 2004] with the following one:

$$g(t-t_i, x-x_i, y-y_i, m_i) = \frac{K*e^{a*(m_i-M_0)}*e^{-t/\tau}}{\{t-t_i+c\}^{1+\omega}*\{(x-x_i)^2+(y-y_i)^2+d(x,y)*e^{\gamma*(m_i-M_0)}\}^{1+\rho}} \quad (S1)$$

Note that an advantage of using and exponentially tapered Omori kernel instead of pure Omori kernel is that the time exponent, $\omega$, is not mathematically constrained to be positive. We then use the Expectation-Maximization scheme proposed by Veen and Schoenberg [2008] for the inversion of ETAS parameters ($K, a, c, \omega, \tau, d, \gamma, \rho$). Table S1 shows the results of inversion for both catalogs, using triggering kernels with pure and exponentially tapered Omori laws. We find that the values of $\tau$ inverted for YANG and HAUK catalogs are respectively ~295 days and ~346 days. They are in both cases in good agreement with the characteristic times inverted from the waiting time distribution, which varies between ~95 and ~180 days. Also, the consistency of the values of $\tau$ between both catalogs demonstrates that $\tau$ remains unaffected by the choices made in the extraction of the YANG catalog from the HAUK catalog. We also note that, for a given catalog, while all the other parameters seem to be in agreement between the pure Omori and tapered Omori cases, c and $\omega$ show significant changes. Both c and $\omega$ seem significantly smaller in the case of the tapered Omori case. Larger values of c and $\omega$ in the pure Omori case is due to the ETAS model trying to fit a genuine tapered Omori decay in the real data with a pure Omori decay. This naturally leads to an overestimation of the c and $\omega$ values.

**Text S4: Comparison of the quality of fits of the inverted waiting time distributions for all the stress bins and the four cases (PREF+, PREF-, RAND+, RAND-)**

In supplementary Figure S13a, we compare the relative goodness of the fit of the inverted models of the waiting time distribution for all stress bins and the four cases shown in the legend, using the measure of maximum log likelihood per data point (MLL) defined in section 3.3.1 (also see Table 1). We find that MLL increases with increasing $|\Delta CFS_{ij}|_k^{median}$ for both positive and negative stress bins. This observation holds for both PREF and RAND. Further, we observe that $\text{MLL}^{k+}$ seems to be larger than the corresponding $MLL^{k-}$ for both PREF and RAND. These observations could be the result of a change of the parameters of the waiting time distributions, a degradation of the fit due to a less appropriate model, or a combination of both. However, in Figure S13b, we demonstrate that the quality of fit of the inverted models for all the stress bins is nearly constant. To show this, we first fit an observed waiting time distribution and compute its maximum likelihood per point (hereafter coined $ML_{obs}$). We then use the parametric analytical form of the inverted distribution to generate a synthetic dataset with the same number of points as the natural one. We then estimate the likelihood per point, $ML_{syn}$, of the synthetic waiting time distribution given the inverted model. This represents the optimal log-likelihood that could be observed for that model if it were the true model. We then compare $ML_{obs}$ to $ML_{syn}$ by computing their ratio. If the model is a correct one for the natural data, the ratio should be close to 1. If the model is bad, the ratio should drastically decrease. We find that this ratio is independent of the amplitude of the Coulomb stress change for all the four cases, and that its value is very close to 1. These observations suggest that the tapered Omori model we used to fit the data is a very good description for all cases and stress bins considered. The apparent change of likelihood with stress level shown on Figure S13a is thus only due changing fitting parameters, and not to variations of the adequacy of the model itself. The fact that the fit quality is also very good for the RAND cases stems from the fact that the model explains both PREF+ and PREF- data. Mixing them due to stress uncertainties thus doesn't change the quality of the fit, even for a more unrealistic model such as RAND, or even our modified, exponentially tapered ETAS model (which is isotropic in nature).

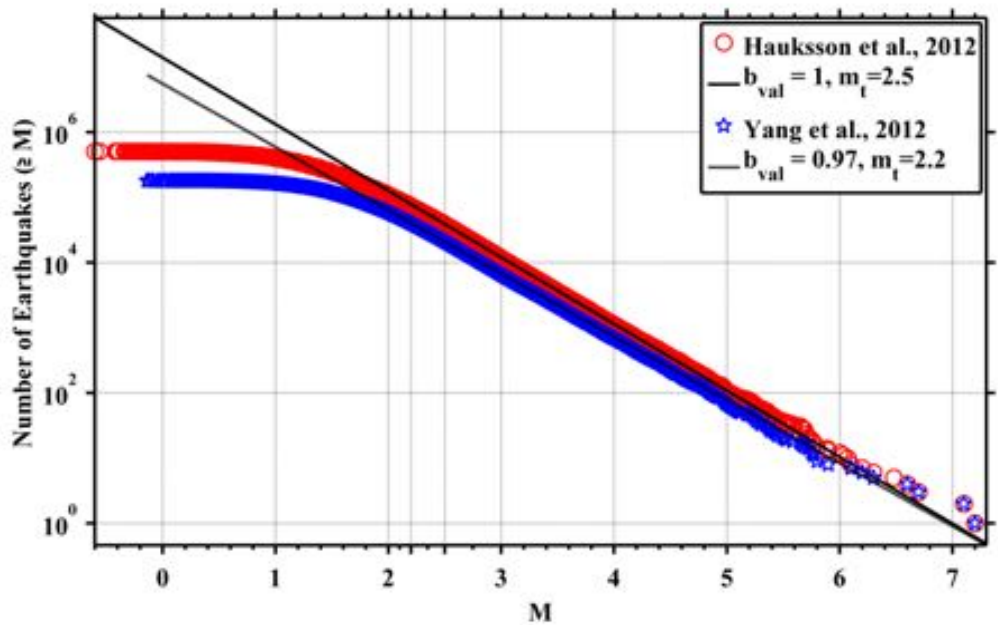

**Figure S1**: Empirical frequency-magnitude distribution for relocated catalog of Southern California from Hauksson et al. [2012] (red circles) and focal mechanism catalog of Southern California from Yang et al. [2012] (blue stars); Solid and dashed lines show the best fit Gutenberg Richter law for the two catalogs above $m_t$ of 2.5 and 2.2 respectively; the legend shows $b_{val}$ corresponding to both catalogs; Both $m_t$ and $b_{val}$ have been inferred using Clauset et al. [2009] method for both catalogs.

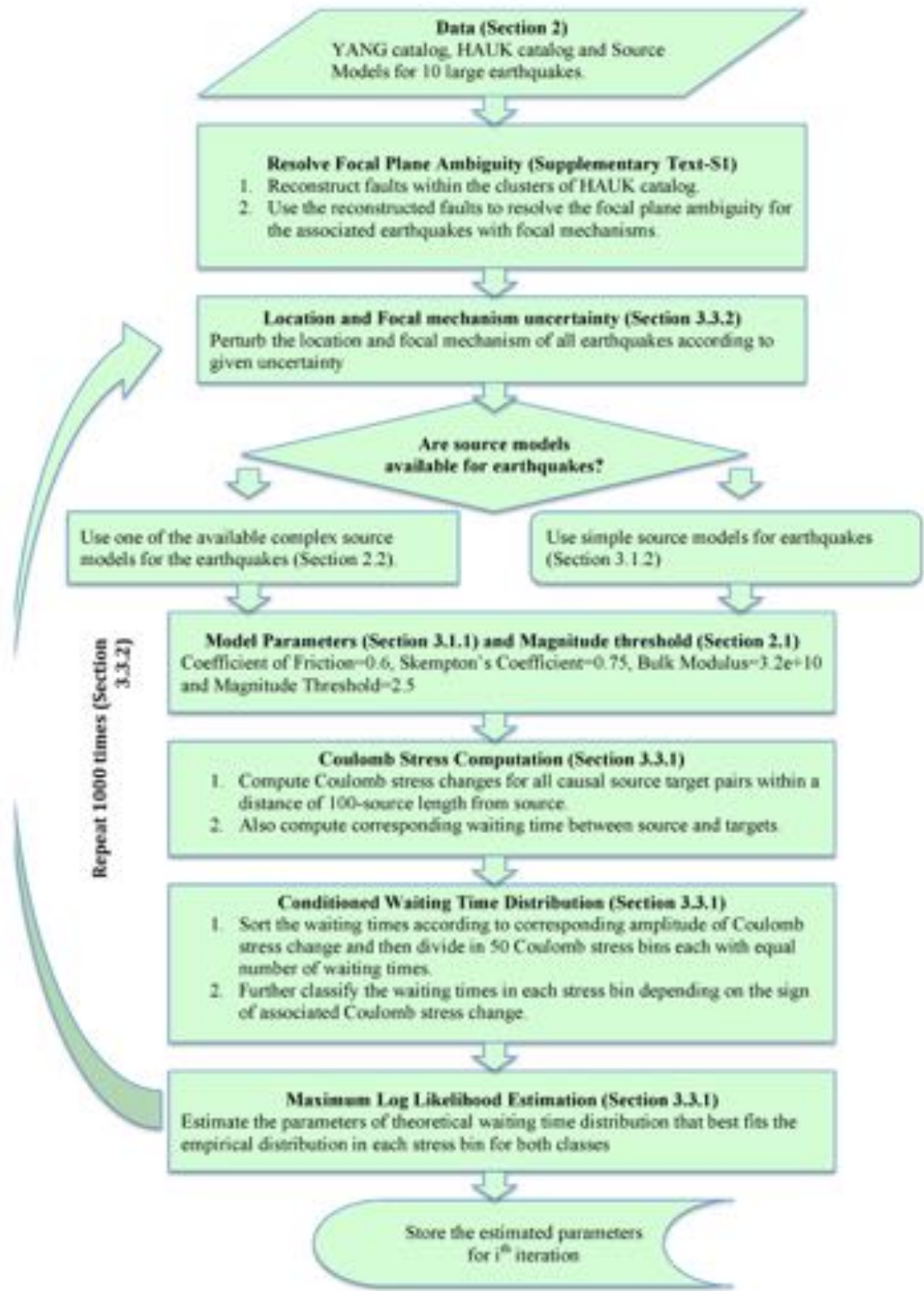

**Figure S2**: Flowchart of our general methodology for the estimation of the waiting time distribution parameters.

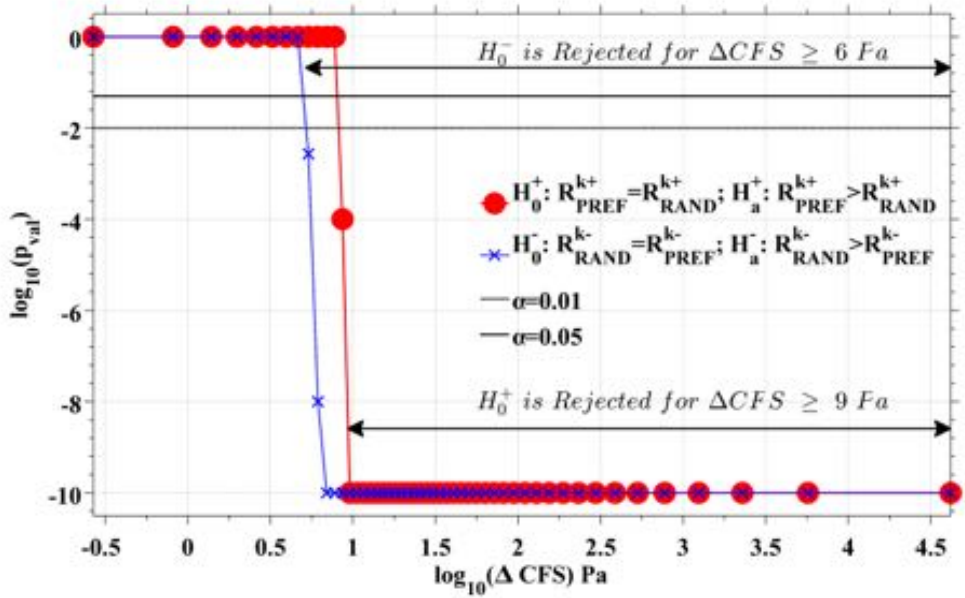

**Figure S3**: $p_{val}$ of Ranksum test vs. Coulomb stress change for positive Coulomb stress bins (Red circles) and negative Coulomb stress bins (Blue crosses); the null hypothesis and the alternative tested in both cases are shown in the legend; the significance level of 0.01 and 0.05 are shown in black dashed line and solid line respectively; the Coulomb stress change thresholds above which the null hypothesis can be rejected in both cases are above ~9 Pa and ~6 Pa respectively.

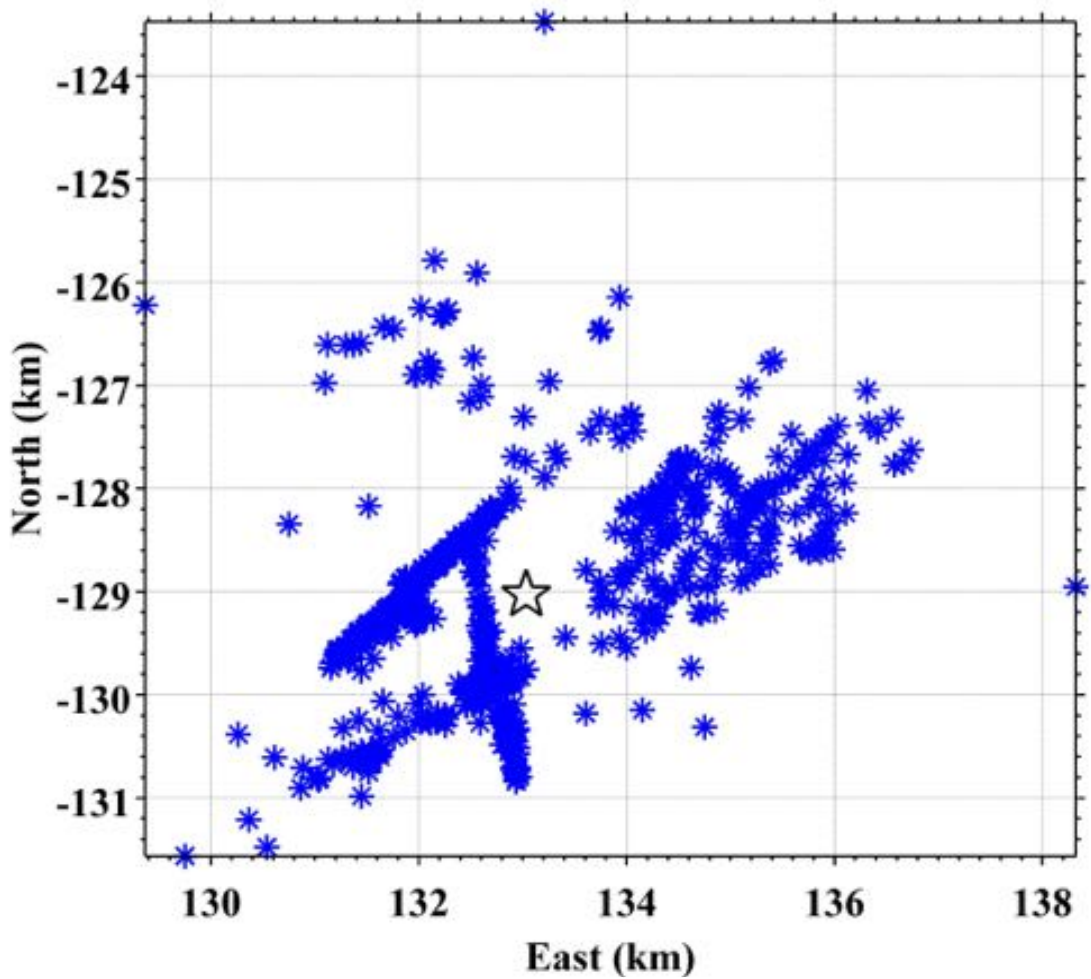

**Figure S4**: Spatial distribution (scales in km) of epicenters of earthquakes belonging to cluster #50106; the geographical coordinates of the barycenter (−115.6143°, 33.0961°) of the cluster is indicated using a black star for reference; the epicenters clearly indicate the presence of multiple faults and spatially uncorrelated events.

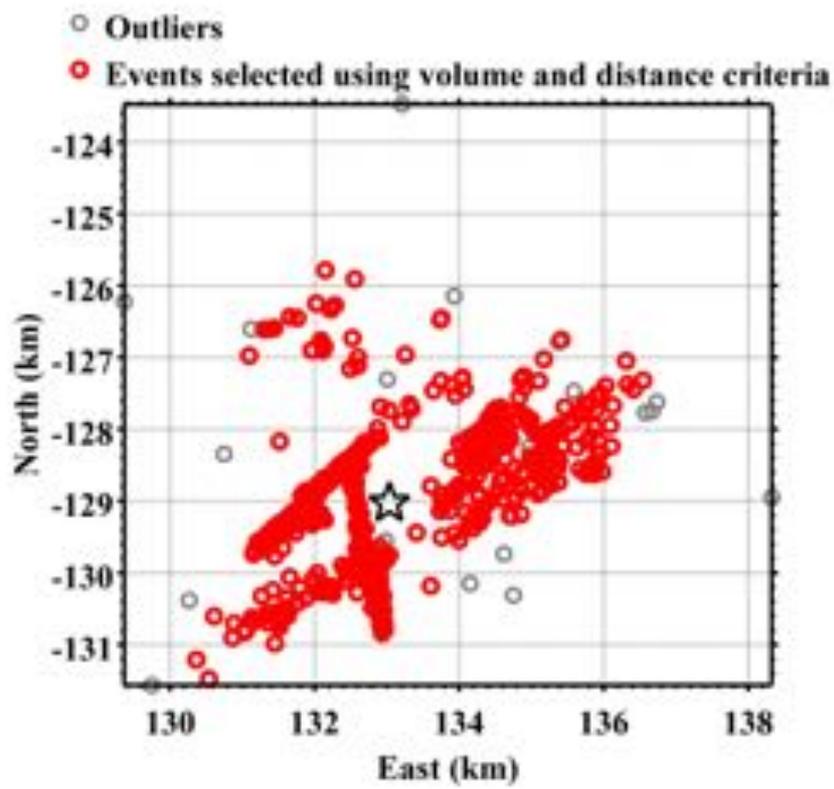
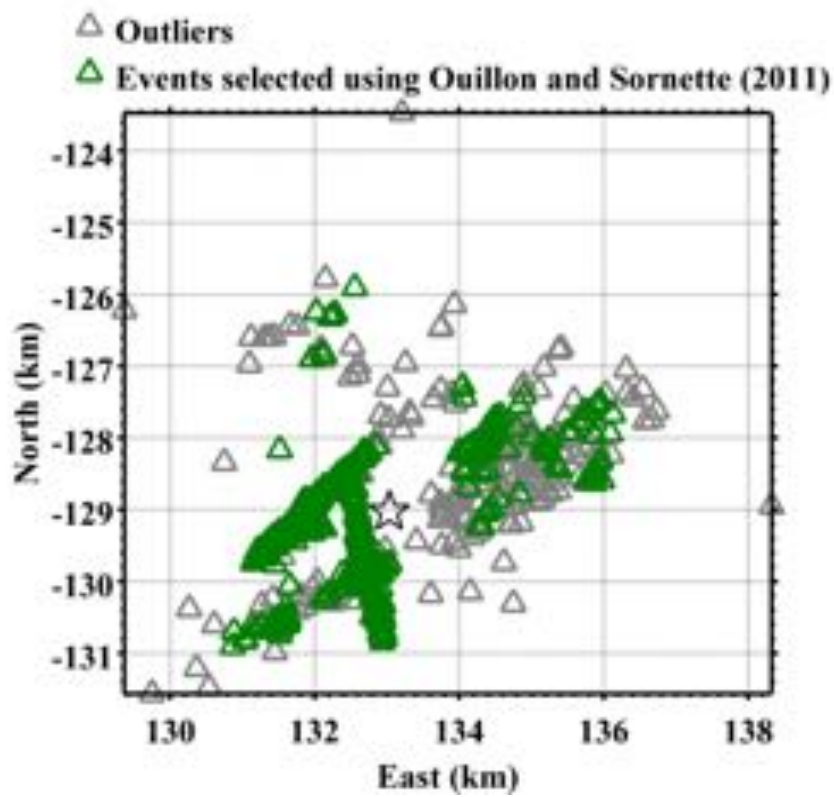

**Figure S5**: **(a)** Epicenters of earthquakes belonging to cluster #50106 left after application of filtering method F2, based on volume and distance criteria (red circles); grey circles stand for events defined as outliers; **(b)** same, using filtering method F1, proposed by Ouillon and Sornette [2011] using green triangles for clustered events; grey triangles are defined as outliers; The geographical coordinates of the barycenter ($-115.6143°, 33.0961°$) of the cluster is indicated using black star for reference.

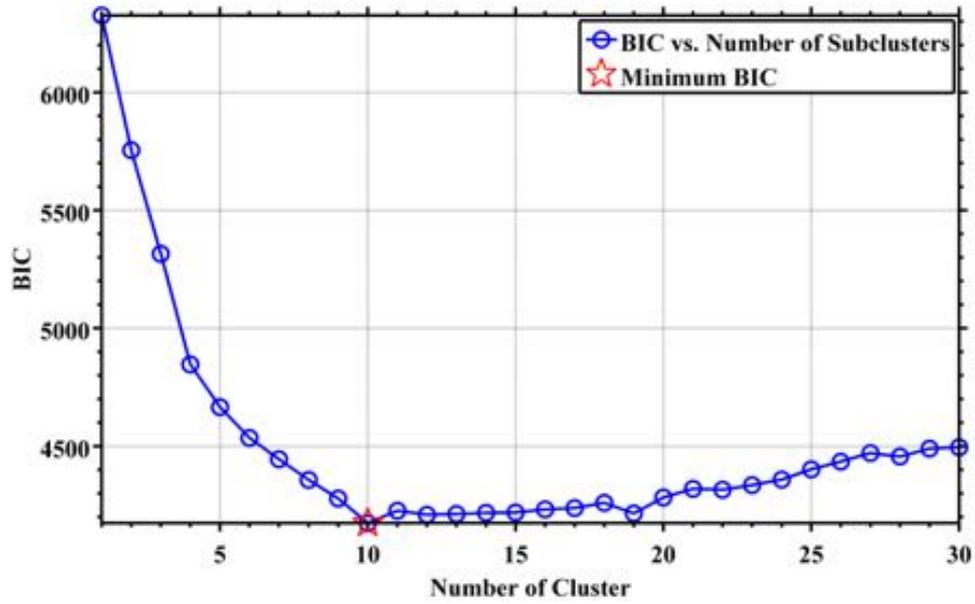

**Figure S6**: BIC of optimal partitioning of the correlated earthquakes in cluster #50106 (Figure S5a, Red Circles) as a function of the number of sub-clusters (Blue Circles); The minimum BIC is indicated by the red star.

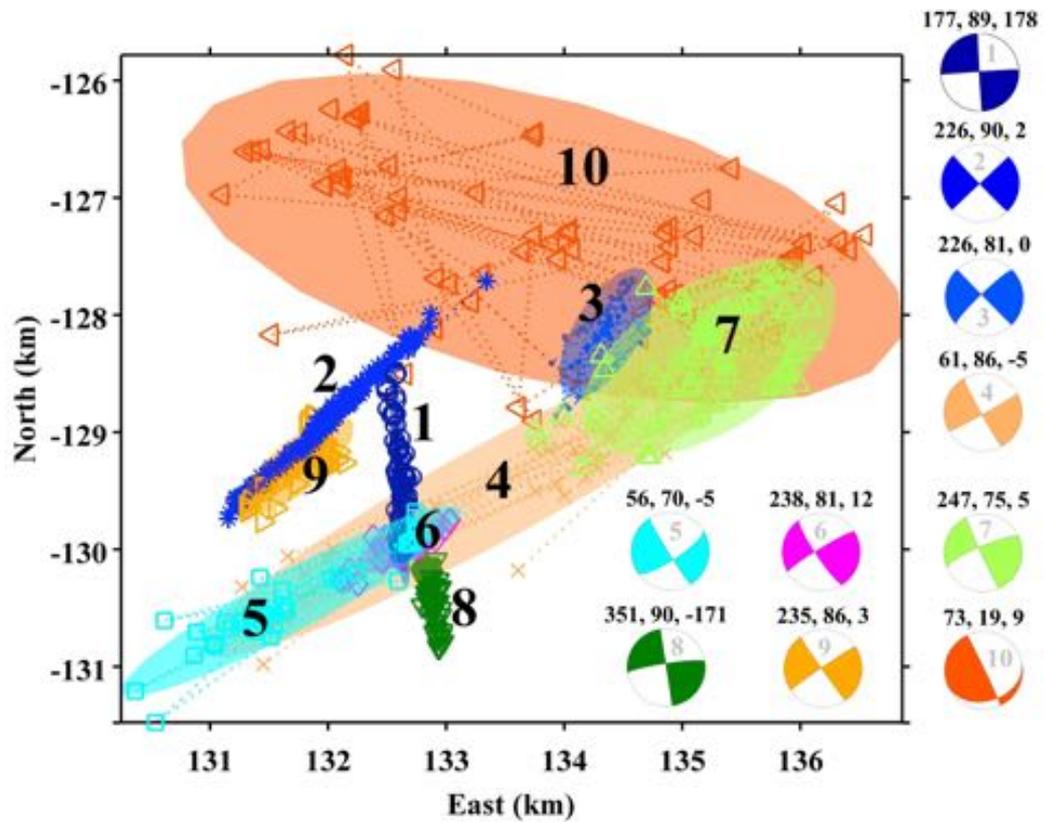

**Figure S7**: Classification of correlated seismicity (Figure S5a, Red Circles), obtained after application of filtering method F2, present in cluster #50106 in 10 Gaussian sub-clusters; the figure shows the horizontal cross-sections of the ellipsoids; Epicenters of earthquakes assigned to each sub-cluster are shown using different marker types and colors (same as parent ellipsoid); Composite focal mechanisms associated with each sub-cluster are shown using colored beach balls with the same color as the associated ellipsoid, and have been assigned same index (shown using bold grey numbers) as that of the corresponding sub-cluster (shown using bold black numbers); strike, dip and rake of the preferred fault plane, obtained using the method proposed in Supplementary Text S1.3, is shown above beach balls for each sub-cluster.

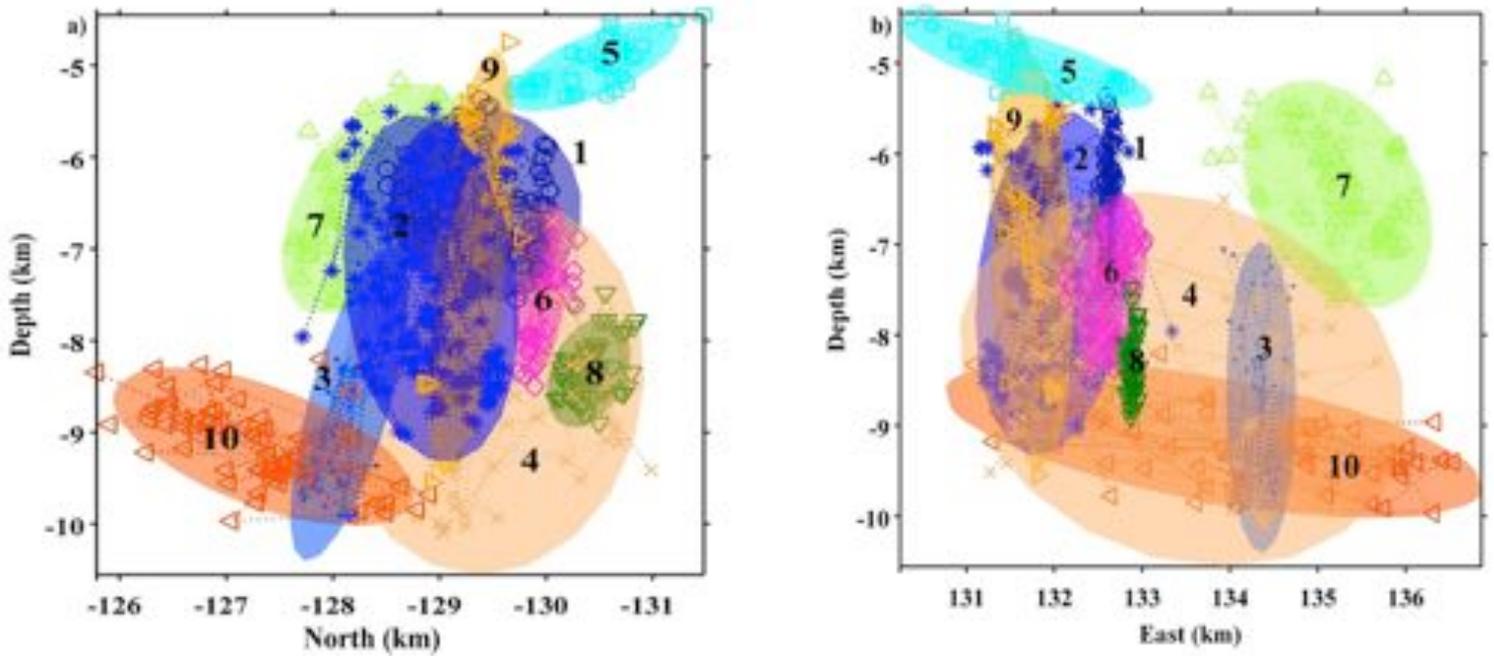

**Figure S8**: Classification of correlated seismicity **(**Figure S5a, Red Circles**)** present in cluster #50106 in 10 Gaussian sub-clusters; the figure shows the two vertical, **(a)** Depth vs. North and **(b)** Depth vs. East cross-sections of ellipsoids; the earthquakes assigned to each sub-cluster are shown using different marker types and colors (same as parent ellipsoid); The indices of the faults correspond to the indices shown in Figure S7.

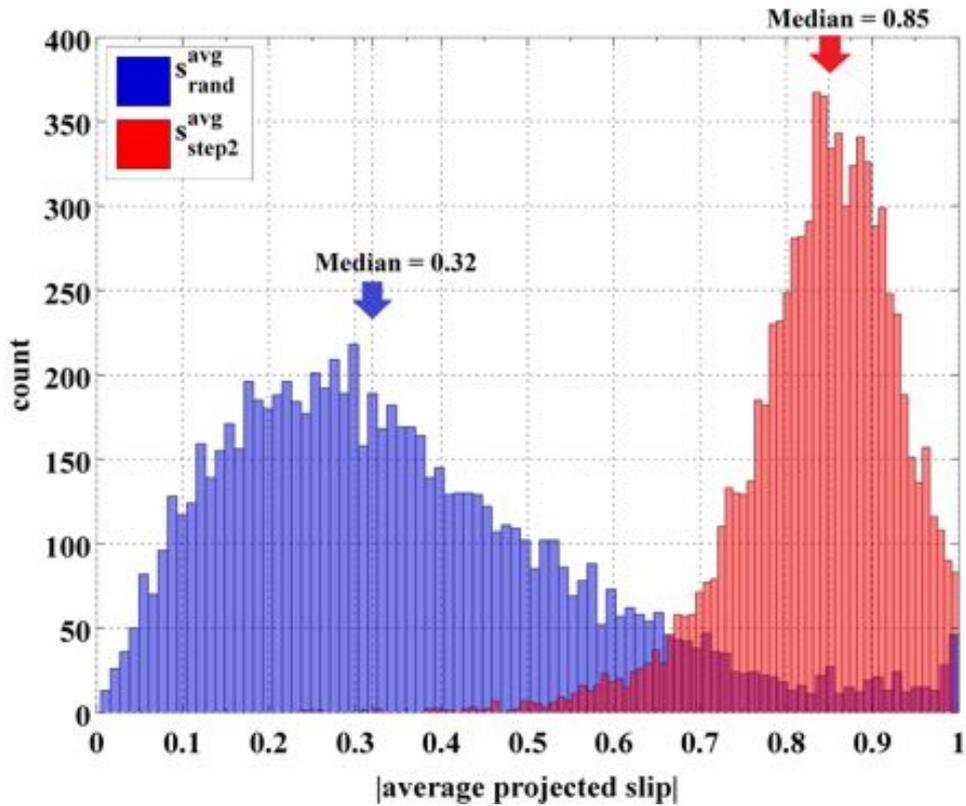

**Figure S9**: Distribution of the amplitudes of the average projected slips on the reconstructed plane of the cluster for the preferred choice of nodal plane ($s^{avg}_{step2}$, shown in red) and random choice of nodal plane ($s^{avg}_{rand}$, shown in blue); Median value for each case in indicated by the solid arrows (same color as the histogram); On average, choosing the nodal planes according to the prescribed method leads to more tightly oriented slip vectors compared to the random choice method.

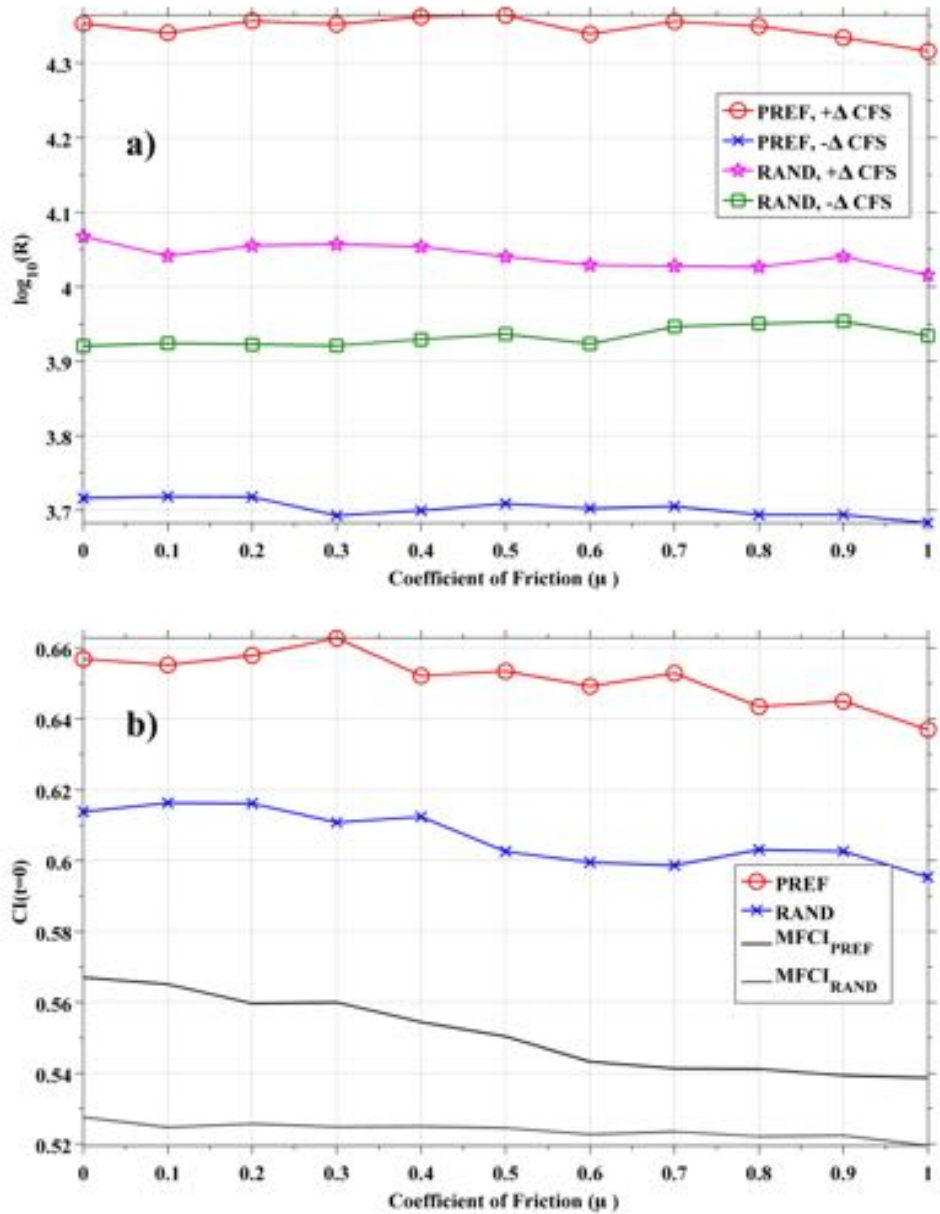

**Figure S10**: **(a)** R vs. $\mu$ (Coefficient of friction) for the largest Coulomb stress bin with a median Coulomb stress change of 42 kPa; Legend shows the labels corresponding to the four cases plotted using different markers **(b)** $CI\ (t=0)$ & $MFCI$ vs. $\mu$ for both PREF and RAND for the largest Coulomb stress bin with median Coulomb stress change of 42 kPa.

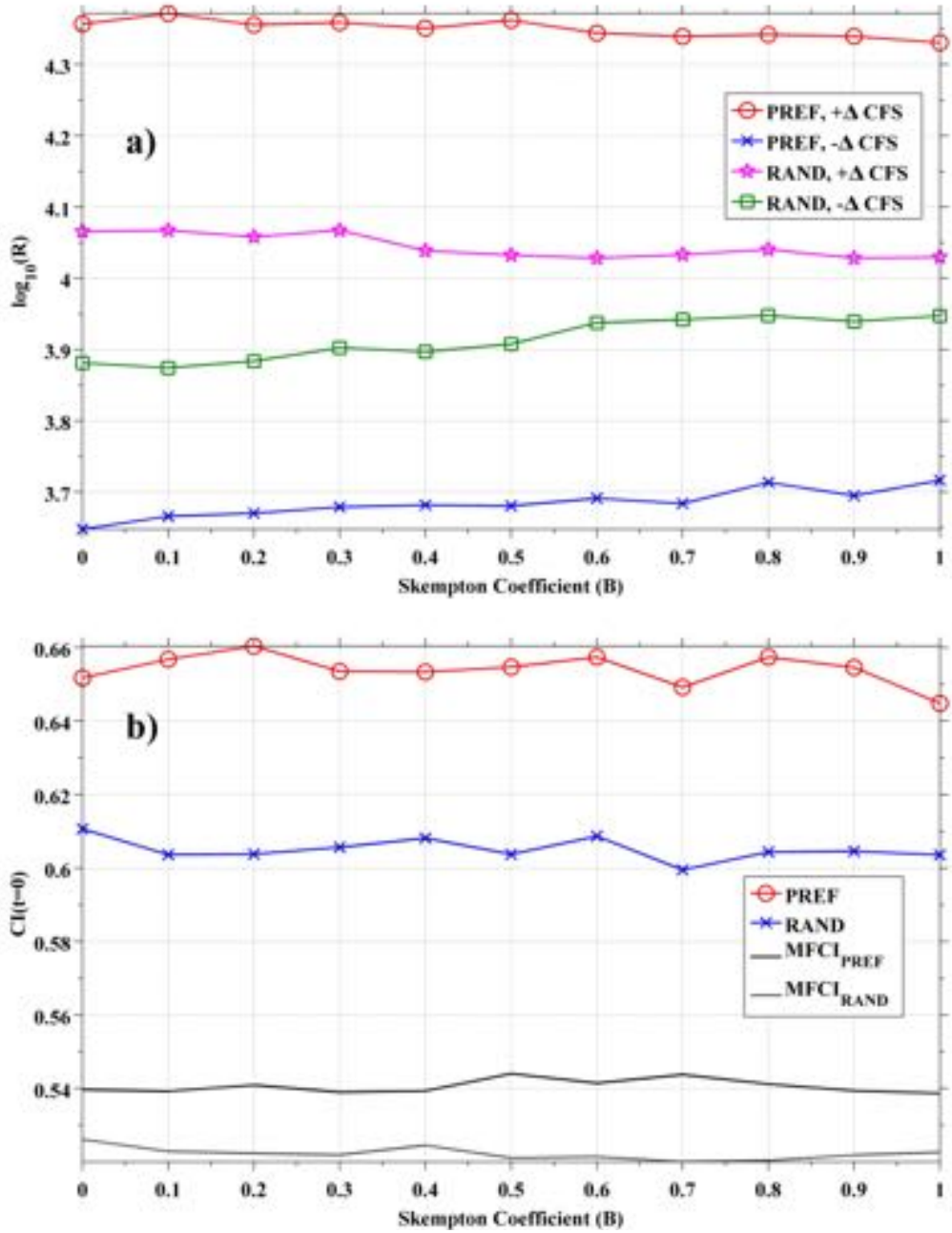

**Figure S11**: **(a)** R vs. $B$ (Skempton's Coefficient) for largest Coulomb stress bin with median Coulomb stress change of 42 kPa; Legend shows the labels corresponding to the four cases plotted using different markers **(b)** $CI\ (t=0)$ & $MFCI$ vs. $B$ for both PREF and RAND for the same stress bin.

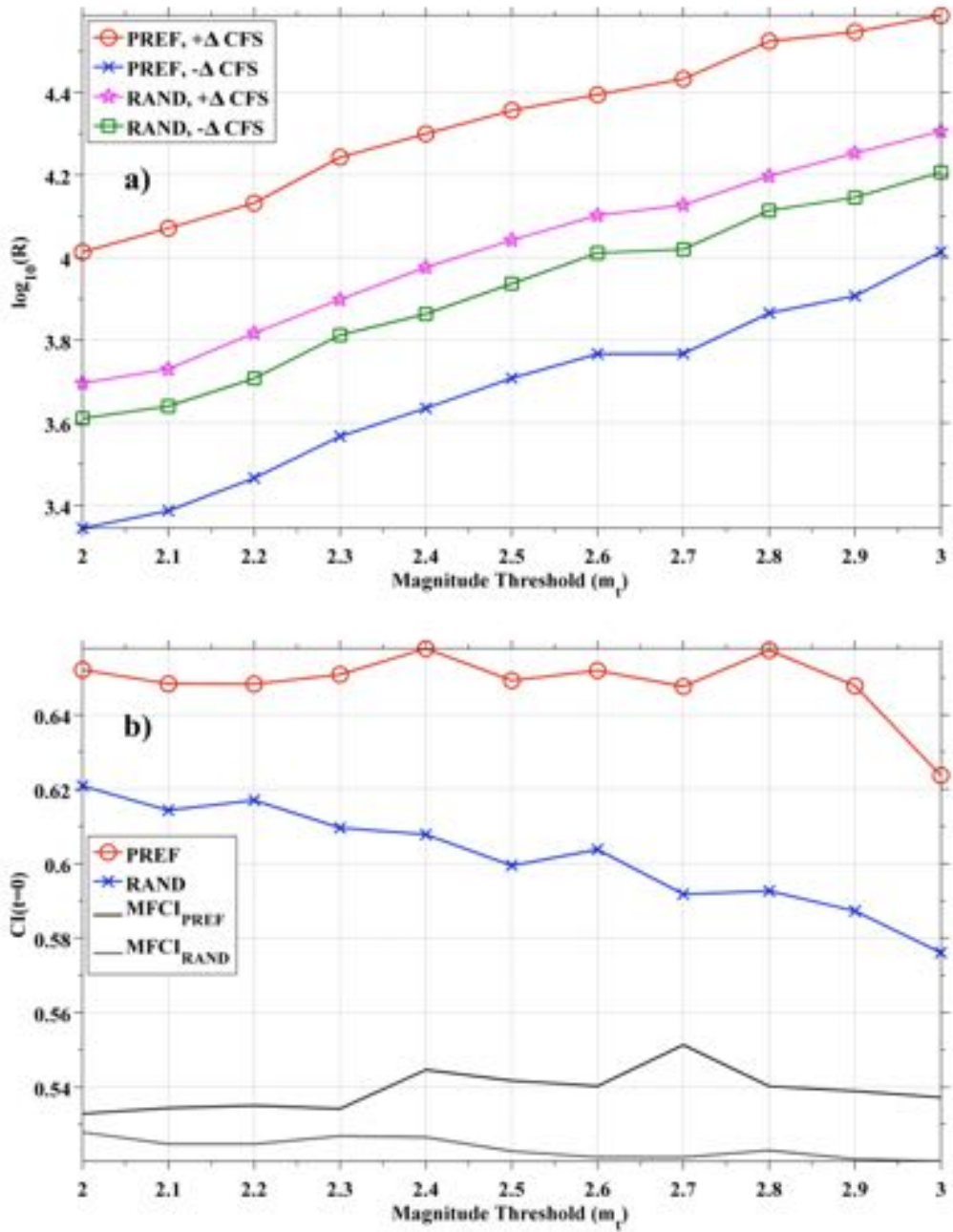

**Figure S12**: **(a)** R vs. $m_t$ (Magnitude Threshold) for largest Coulomb stress bin with median Coulomb stress change of 42 kPa; Legend shows the labels corresponding to the four cases plotted using different markers **(b)** $CI\ (t=0)$ & $MFCI$ vs. $m_t$ for both PREF and RAND for the same stress bin.

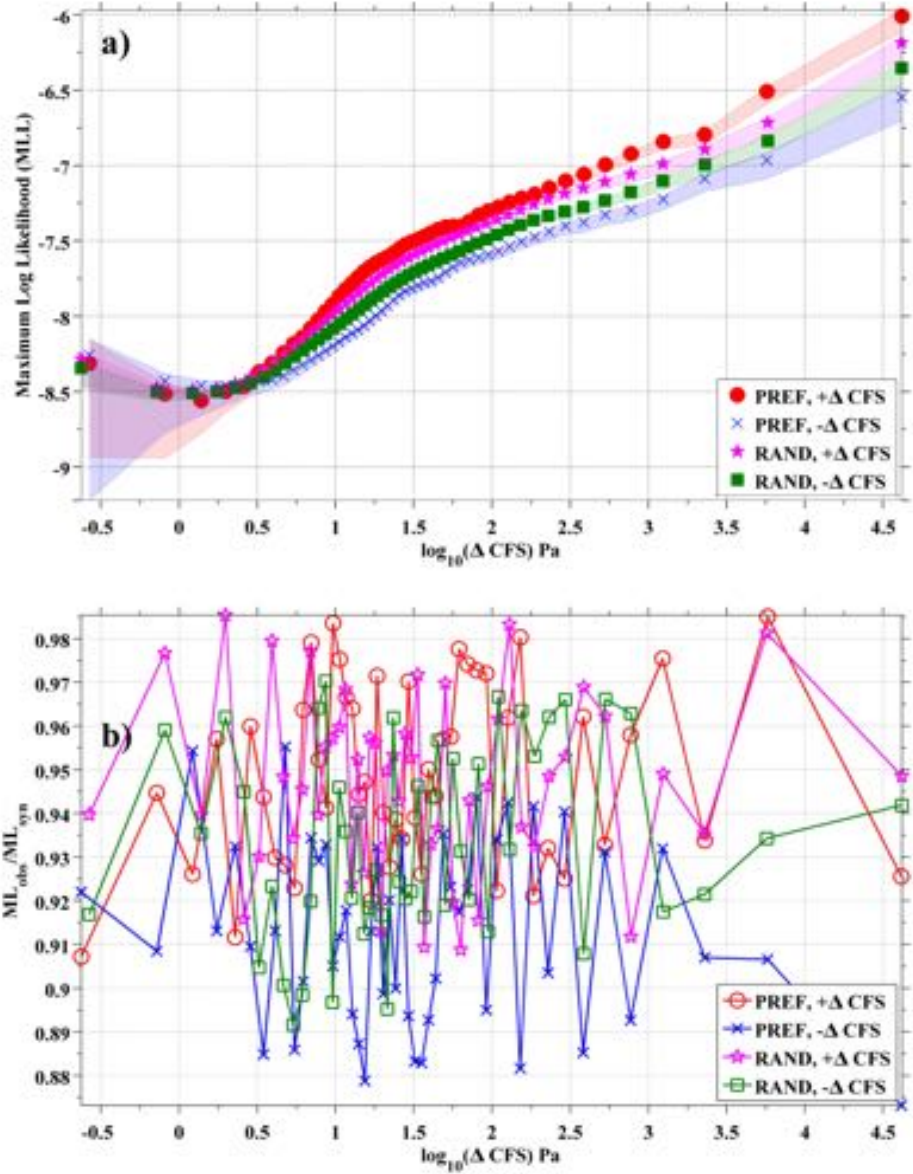

**Figure S13**: **(a)** $MLL^{k+}$ and $MLL^{k-}$ (maximum log-likelihood per point) vs. Coulomb stress change for both PREF and RAND; The legend shows the labels corresponding to the four cases; the median value corresponding to each case is plotted using markers; the shaded region plotted in the same colors as the markers shows the 95% confidence interval delineated by the 2.5% and 97.5% quantiles; **(b)** Ratio of observed maximum likelihood per point, $ML_{obs}$, and likelihood of synthetic data generated using the inverted models, $ML_{syn}$, as function of Coulomb stress change; the legend shows the labels corresponding to the four cases; the median value corresponding to each case is plotted using markers.

**Table S1**: Inverted parameters of the ETAS model with pure Omori and exponentially tapered Omori decay for YANG and HAUK catalogs.

|  |  | K | $\alpha$ | $c$ (days) | $\omega$ | $\tau$ (days) | $d$ ($km^2$) | $\rho$ | $\gamma$ |
|---|---|---|---|---|---|---|---|---|---|
| YANG | Pure Omori | 0.0021 | 1.75 | 0.077 | 0.29 | NA | 0.030 | 0.50 | 1.48 |
|  | Tapered Omori | 0.0013 | 1.70 | 0.0037 | -0.068 | 295 | 0.029 | 0.48 | 1.49 |
| HAUK | Pure Omori | 0.0040 | 1.57 | 0.033 | 0.26 | NA | 0.19 | 0.50 | 1.10 |
|  | Tapered Omori | 0.0030 | 1.53 | 0.0018 | -0.064 | 346 | 0.19 | 0.49 | 1.10 |

## Bibliography for Supporting Information